\documentclass[aps,prd,preprint,preprintnumbers,nofootinbib]{revtex4}
\usepackage{graphicx,epsfig,xcolor}
\usepackage{latexsym,amsmath,amssymb,amstext}
\usepackage{float}
\usepackage{longtable}
 \usepackage{ulem} 
\usepackage[paperwidth=210mm,paperheight=297mm,centering,hmargin=2cm,vmargin=2.5cm]{geometry}

\newcommand{\be}{\begin{equation}}
\newcommand{\ee}{\end{equation}}
\newcommand{\bea}{\begin{eqnarray}}
\newcommand{\eea}{\end{eqnarray}}
\newcommand{\Tr}{{\rm Tr}}

\newcommand\bef{\begin{figure}}
\newcommand\eef[1]{\label{fg:#1}\end{figure}}
\newcommand\beq{\begin{equation}}
\newcommand\eeq[1]{\label{#1}\end{equation}}
\newcommand\beqa{\begin{eqnarray}}
\newcommand\eeqa[1]{\label{#1}\end{eqnarray}}
\newcommand\bet{\begin{table}}
\newcommand\eet[1]{\label{tb:#1}\end{table}}

\newcommand\fgn[1]{Figure \ref{fg:#1}}
\newcommand\eqn[1]{Eq.\ (\ref{#1})}

\newcommand\apx[1]{Appendix \ref{sec:#1}}
\newcommand\tbn[1]{Table \ref{tb:#1}}

\newcommand\ie{{\sl i.e.\/}}

\begin{document}

\title{Scale-invariance and scale-breaking in parity-invariant three-dimensional QCD}
\author{Nikhil\ \surname{Karthik}}
\email{nkarthik@bnl.gov}
\affiliation{Physics Department, Brookhaven National Laboratory, Upton, New York 11973-5000, USA}
\affiliation{Department of Physics, Florida International University, Miami, FL 33199}
\author{Rajamani\ \surname{Narayanan}}
\email{rajamani.narayanan@fiu.edu}
\affiliation{Department of Physics, Florida International University, Miami, FL 33199}

\begin{abstract}
We present a numerical study of three-dimensional two-color QCD
with $N=0, 2, 4, 8$ and 12 flavors of massless two-component fermions
using parity-preserving improved Wilson-Dirac fermions.  A finite
volume analysis provides strong evidence for the presence of
$Sp(N)$ symmetry-breaking bilinear condensate when $N\le2$
and its absence for $N\ge8$. A weaker evidence for the bilinear
condensate is shown for $N=4$.  We estimate the critical number of
flavors below which scale-invariance is broken by the bilinear
condensate to be between $N=4$ and 6.
\end{abstract}

\date{\today}
\maketitle

\section{Introduction}

Three-dimensional gauge theories coupled to an even number of
two-component massless fermions can be regularized to form a
parity-invariant theory. The parity-invariant
fermion action for $N$ flavors of two-component fermions coupled
to $SU(N_c)$ gauge-field $A_\mu$ is
\be
S_f = \int d^3x\sum_{i=1}^{N/2}\left\{\overline\phi_i(x) C(A) \phi_i(x) + \overline\chi_i(x) C^\dagger(A) \chi_i(x)\right\},
\ee
where $C(A)$ is the two-component Dirac operator, with an ultraviolet
regularization being imposed implicitly.  In the continuum,
\be
C(A)=\sum_{\mu=1}^3 \sigma_\mu(\partial_\mu+i A_\mu(x)),
\ee
with $\sigma_\mu$ being the three Pauli matrices. One can think of the above
parity-invariant theory of $N$ flavors of two-component fermions
to be equivalent to a theory of $\frac{N}{2}$ flavors of four-component
fermions $\psi_i$ with a Hermitian Dirac operator $D$:
\be
S_f=\sum_{i=1}^{N/2}\int d^3x\overline{\psi}_i(x) D \psi_i(x),
\ee
with the following identifications
\be
\psi_i(x) =\begin{pmatrix}\phi_i(x)\\\chi_i(x)\end{pmatrix}; \quad \overline\psi_i(x) =(\overline\chi_i(x),\overline\phi_i(x));\quad D(A)=\begin{pmatrix} 
0 & C^\dagger(A) \cr C(A) & 0 \end{pmatrix}.
\label{hermd}
\ee
Since $C^\dagger=-C$ in the continuum, the theory has a global $U(N)$
flavor symmetry. However, the $N_c=2$ theory is special, and there 
is a larger $Sp(N)$ 
global symmetry following from the property 
$\sigma_2\tau_2 C\sigma_2\tau_2=C^t$, where $\sigma_2$ and $\tau_2$ are 
Pauli matrices in spin and color space respectively~\cite{Magnea:1999iv}.
This is also related to the fact that the operator $D$ can be made real 
symmetric in a suitable basis~\cite{Neuberger:1998wg}.

Consider such a theory on an Euclidean periodic $l_{\rm ph}^3$
torus. Physics depends on the dimensionless size, $\ell = l_{\rm
ph} g^2_{\rm ph}$, where $g^2_{\rm ph}$ is the physical coupling
constant. Since $g^2_{\rm ph}$ has the dimension of mass, these
theories are super-renormalizable, and the continuum limit in a
lattice regularization at a fixed $\ell$ can be obtained by setting
the lattice coupling constant (same as the lattice spacing) to
$g^2_{\rm lat} = \frac{\ell}{L}$ on a periodic $L^3$ lattice and
taking $L\to\infty$. The physics of this theory will smoothly
cross-over from a non-interacting theory at small $\ell$ to a
strongly interacting theory as $\ell\to\infty$; this strongly
interacting theory could either be scale-invariant or scale-breaking.
Scale-breaking is expected to produce a parity-preserving non-zero
fermion bilinear condensate, that for a generic $SU(N_c)$ gauge
theory breaks the $U(N)$ flavor symmetry to $U\left(\frac{N}{2}\right)
\times U\left(\frac{N}{2}\right)$~\cite{Pisarski:1984dj,Vafa:1984xh},
but in the case of $SU(2)$ gauge theory breaks the $Sp(N)$ symmetry to
$Sp\left(\frac{N}{2}\right)\times
Sp\left(\frac{N}{2}\right)$~\cite{Magnea:1999iv}.  The possibility
of the generation of any parity-breaking bilinear condensate has
been argued against in~\cite{Vafa:1984xh}.

A numerical study of the Abelian $U(1)$ gauge theory using Wilson
fermions showed no evidence for a bilinear condensate for any even
value of $N>0$~\cite{Karthik:2015sgq}.  Since the Wilson fermions
realize the $U(N)$ flavor symmetry only in the continuum limit,
overlap fermions were used to study the theory with the symmetry
present even away from the continuum. This enabled us to study the
scale-invariant properties of the theory in further
detail~\cite{Karthik:2016ppr}. Assuming the $N=2$ theory to be
scale-invariant, a strong infra-red duality~\cite{Wang:2017txt,Qin:2017cqw}
predicts an enhanced $O(4)$ symmetry and this was also verified
using overlap fermions~\cite{Karthik:2017hol}. Since the Abelian
theory is scale invariant for all even values of $N$, we turn to
$SU(N_c)$  non-Abelian theory in this paper in order to study a
transition from a scale invariant theory to one that breaks scale
invariance as one changes the number of flavors. It is worth noting that three dimensional
$SU(2)$ gauge theories with even number of massless fermions appear
in the study of spin
liquids~\cite{Wen:2004ym,Lee:2006zzc,PhysRevB78054432,Wang:2017txt}.  The
non-Abelian theory in the 't Hooft limit (number of colors going
to infinity at a fixed number of flavors) has been shown to have a
non-zero bilinear condensate~\cite{Karthik:2016bmf}.  Dagotto,
Koci\'c and Kogut~\cite{DAGOTTO1991498} numerically investigated
the $SU(2)$ theory with staggered fermions on small lattices (mainly
on $L=8$).  They tried to deduce the presence or absence of the
condensate in the massless limit by computing it at two different
fermion masses and then extrapolating it to zero.

The four-dimensional $SU(N_c)$ gauge theory coupled to massless
fermions becomes infra-red free and loses asymptotic freedom above
a certain number of fermion flavors. In $4-\epsilon$ dimensions,
$\epsilon$-expansion calculation suggests the infra-red free behavior
develops into a non-trivial conformal (IR) fixed point if $N >
11N_c$~\cite{Peskin:1995ev,Goldman:2016wwk}.  Also, an earlier
analysis using the Schwinger-Dyson equations~\cite{Appelquist:1989tc}
using $1/N$-expansion in three dimensions, similar to the one for
the Abelian theory~\cite{Appelquist:1988sr}, suggests that the
theory is scale invariant if $N >\frac{256}{3\pi^2}\frac{N_c^2
-1}{N_c}$.  However, a study of the flow of four-fermion operators
in the $\epsilon$-expansion~\cite{Goldman:2016wwk} suggests a new
IR fixed point, different from the one at large-$N$, might be present
if
$
N < 11 N_c +\left[ 12+\frac{8}{N_c} + O\left(N_c^{-2}\right) \right]\epsilon + O(\epsilon^2).
$
It is tempting to identify the lower bounds on $N$ as a critical
$N^*$ below which scale invariance is broken but one has to consider
the possibility that it is a point that separated one class of IR
fixed points from another.  There is indication of such a scenario
in QED$_3$ where the IR fixed point at $N=2$ has an $O(4)$ symmetry
unlike the IR fixed point at $N=\infty$. Therefore, a first principle
numerical estimate of an $N^*$ that is below any of the bounds
obtained from other calculations ( with their own underlying
assumptions)  has interesting implications.

We are, therefore, motivated to perform a careful numerical study
of the $N_c=2$ theory with $N$ flavors of massless two-component
fermions with the sole aim of obtaining the critical number of
fermions, $N^*$, such that the theory develops a scale for $N <
N^*$ and is scale-invariant for $N \ge N^*$.  Since we are interested
in numerically studying several different values of $N$ and our aim
is only to locate the critical number of flavors, we use improved
Wilson fermions~\cite{Karthik:2015sgq} in order to reduce the
computational cost. A careful study of the transition from above
$N^*$ to below $N^*$ will require overlap fermions~\cite{Karthik:2016ppr}
or domain wall fermions~\cite{Hands:2015dyp,Hands:2015qha}.  The
type of lattice fermions one uses is irrelevant for the continuum
physics; however, the main difference arises at finite lattice
spacings where the propagator of two-component Wilson fermions is
not anti-Hermitian but the propagator of two-component overlap
fermions is anti-Hermitian.  We reserve the computation of using
$U(N)$ symmetry preserving lattice fermions for the future, which
will give us more information on the fermion bilinear correlators
and serve the purpose of cross-checking the continuum results in
this study.

\section{Finite volume analysis of low lying eigenvalues of the massless Dirac operator}

The basic philosophy of the finite volume analysis in this paper
is the same as in~\cite{Karthik:2015sgq}.  The eigenvalues of the
Hermitian Dirac operator $D$ in \eqn{hermd} occur in positive-negative
pairs $\pm \lambda$ given by the equation
\be
C^\dagger C u_\lambda = \lambda^2 u_\lambda.
\ee
The eigenvalues $\lambda$ are gauge invariant and discrete at finite
$\ell$.  Therefore, we study the ordered discrete spectrum of
eigenvalues of $\sqrt{C^\dagger C}$:
\be
0 < \lambda_1(\ell) < \lambda_2(\ell) < \cdots,
\ee
obtained as observables in the $N$ flavor theory of massless fermions.
Henceforth, we will use $\lambda_i$ to denote the expectation value
of the $i$-th lowest eigenvalue of $\sqrt{C^\dagger C}$ over the
gauge-field path integral.  
The asymptotic behavior of $\lambda_i(\ell)$ as $\ell\to\infty$
falls into three types:
\begin{itemize}

\item[Type 1:] Free field behavior will result in $\lambda_i(\ell)
\propto \frac{1}{\ell}$ with the proportionality constants determined
by the mean value of the gauge field which is equivalent to the
induced boundary conditions on the fermions. Such a behavior is certainly
expected in small $\ell$ due to the asymptotic freedom in the theory.

\item[Type 2:] A complete level repulsion between the eigenvalues
will result in $\lambda_i(\ell) \propto \frac{1}{\ell^3}$ with the
proportionality constants determined by one value of the bilinear
condensate $\Sigma$ that breaks scale-invariance.

\item[Type 3:]  In an interacting scale-invariant theory, the
eigenvalues which have the naive dimension of mass,  will scale as
$\lambda_i(\ell) \propto \frac{1}{\ell^{1+\gamma_m}}$ with $ 0 <
\gamma_m < 2$ being the mass anomalous dimension. Here, we are
assuming there are no further restrictions in three dimensions on
the possible values $\gamma_m$ can take between 0 and 2.

\end{itemize}
Since the theory at very small $\ell$ will be non-interacting, we
expect a smooth cross-over from the $\frac{1}{\ell}$ behavior at
small $\ell$ to one of the above cases as $\ell\to\infty$ \footnote{
One could have taken a different approach and kept the physical
extent in one of the directions fixed at a value $1/T$ while taking
$\ell\to\infty$ in other directions, in which case we would be
studying the theory at finite temperature $T$, and there might  be
singular behavior around some $T_c$.  We are not taking that approach
here (c.f. \cite{Damgaard:1998yv} for such an approach 
in three-dimensional quenched SU$(3)$ theory).}. As an abuse of notation, 
we will use $\gamma_m=2$ to mean the scale-breaking type (2) behavior even though 
the exponent is no longer an anomalous dimension in this case.

We have summarized the numerical details pertaining to the
lattice simulation and the extraction of the low-lying spectrum 
in \apx{numdet}.  
Having chosen a numerical approach to study the theory as a function
of $\ell$, we have to face its limitations.  The lattice spacing
is given by $\frac{\ell}{L}$ and in spite of improving the lattice
action we have to face the fact that as $\ell$ gets large we have
to make $L$ also large to reduce lattice spacing effects. Furthermore,
if $\frac{\ell}{L} > 2.3$ we will be in a lattice theory with strong
coupling and there is a bulk cross-over~\cite{Bursa:2005tk,
Narayanan:2007ug} that separates an unphysical strong coupling phase
from the continuum phase that we want to study.  Using the computing
resources available to us, we were able to go up to $L=28$ in a
theory with $N\ge 2$, and up to $L=32$ in the quenched limit ($N=0$).
If we now ask for acceptable levels of lattice spacing effects and
also require that we are in the continuum phase of the theory we
are led to study the theory for $\ell \le 17$ and this is what we
have performed here.  

At this point, it is appropriate to make a
few remarks about the analysis performed earlier in~\cite{DAGOTTO1991498}.
The lattice coupling constant in~\cite{DAGOTTO1991498} is defined
as $\beta = \frac{4}{g^2_{\rm lat}} = \frac{4L}{\ell}$ and one
should set $\beta > 1.7$ to be in the continuum phase of the theory.
Using such values of $\beta$ which are in the continuum phase,
Dagotto et~al., found indications of non-zero condensate in the
massless theory only for $N=0,2$ and $4$ using $8^3$ lattice by a
linear extrapolation of condensate at two different non-zero fermion
masses. With the increased computational resources available at
present, one could put their observations on a firmer footing by
following a similar approach by simulating a wide range of fermion
masses $m$ at different lattice volumes in order to take a controlled
thermodynamic as well as the massless limits. In this case, one
should also follow an unbiased approach by assuming  a $\Sigma(m)
\sim m^{2-\gamma}+\mathcal{O}(m)$ mass dependence of the condensate
at infinite volume~\cite{DelDebbio:2010ze}, in order to allow for the massless theory to
be scale-invariant.  However, we use the finite-size scaling of
eigenvalues of the massless Dirac operator to determine the presence
or absence of condensate along the lines of our previous studies
of QED$_3$.  This method is also advantageous because we avoid the
presence of two scales $\ell$ and $m$ at once in the problem.

If the theory has a bilinear condensate, the behavior of the
eigenvalues as a function of $\ell$ will smoothly cross-over from
the type (1) to type (2) as we increase $\ell$. The type (2) asymptotic behavior will be given by
\be
\lambda_i(\ell) \sim \frac{z_i}{\Sigma \ell^3}\label{sigdef},
\ee
where $\Sigma$ is the value of the bilinear condensate per fermion
flavor at $\ell=\infty$ and $z_i$'s are universal numbers given by
an appropriate random matrix model~\cite{Magnea:1999iv,Verbaarschot:1994ip}.  If
$1/\sqrt{\Sigma}$ sets the typical spontaneously generated length-scale
in the system, then this cross-over to the asymptotic behavior
happens only for box sizes $\ell \sim
\mathcal{O}\left(1/\sqrt{\Sigma}\right)$, which renders the measurement
of small values of $\Sigma$ computationally difficult.  As we get
closer to the critical number of flavors $N^*$ from below, the value
of the bilinear condensate would get smaller, making it more difficult
to decide if the theory has a non-zero bilinear condensate or not.
If the theory does not have a bilinear condensate the behavior of
$\lambda_i(\ell)$ will smoothly cross-over  from type (1) to type
(3) as $\ell$ is increased and the asymptotic behavior will set in
early in $\ell$ if we are well above $N^*$ since we expect $\gamma_m$
to approach zero (free field behavior) as $N\to\infty$.

With the above picture in mind as we change the number of
flavors, we will start by defining
\be
\Sigma_i(\ell) \equiv \frac{z_i}{\lambda_i(\ell) \ell^3}, \label{sigil}
\ee
where the estimate of $\lambda_i$ is made by averaging over the
eigenvalues measured in different gauge configurations sampled by
Monte Carlo.  We use $z_i$ simply to scale the right-hand side and
we make no assumption about the presence of a non-zero bilinear
condensate; for the type (3) conformal behavior of $\lambda$, $\Sigma(\ell)$ as defined above will 
approach 0 as $\ell^{-2+\gamma_m}$. We fit the data for $\Sigma_i(\ell)$ to the functional form
\be
\Sigma_i(\ell)  = a_0^i(\gamma_m) \ell^{-2+\gamma_m} \left [  1 + \frac{a_1^i(\gamma_m)}{\ell} + \frac{a_2^i(\gamma_m)}{\ell^2}\right]\quad\text{with}\quad a_0^i >0;\quad 0<\gamma_m<2,\label{sigfit}
\ee
where the three fit parameters, $a_k^i(\gamma_m)$, $k=0,1,2$ depend
on the choice of $\gamma_m$ and $i$.  Note that our choice implies
that $a_0^i(\gamma_m)$ cannot be zero implying that the term in
front of the parenthesis is the asymptotic behavior.  By studying
the $\chi^2$ per degree of freedom ($\chi^2/{\rm DOF}$) of the fits
as a function of $\gamma_m$, we will be able to find the value that
best fits the data for a given $i$.  We expect the best value of
$\gamma_m$, where the $\chi^2$ is minimized, to be independent of
$i$.  As per our discussion in the previous paragraph, we expect
this approach to work with relative ease for value of $N$ away from
$N^*$.  Contrary to the form used in \eqn{sigdef}, we can also use
\be
\Sigma_i(\ell)  = \Sigma + \frac{a_1}{\ell} + \frac{a_2}{\ell^2},
\label{sigcond}
\ee
where we are assuming the possibility for a non-zero condensate
$\Sigma$, with the finite volume correction that is Taylor expandable
in $1/\ell$.  This form does not allow for an anomalous dimension,
which is indeed the case when there is a condensate.  Since we will
have results for $\lambda_i(\ell)$ in a finite range of $\ell$, both
\eqn{sigfit} and \eqn{sigcond} should result in the same physical
conclusion well away from $N^*$. But, we expect conflicts between
these two forms closer to $N^*$.  In both the ans\"atze, we could
have included more orders of $1/\ell$ corrections. But empirically,
we find that $1/\ell$ and $1/\ell^2$ corrections are enough to
describe our data well. Therefore, our conclusions have to be
interpreted in a Bayesian sense --- given the priors that only
$1/\ell$ and $1/\ell^2$ finite volume corrections are important in
the data we have, and assuming this continues to be the case at
even larger $\ell$ where we do not have the data, we ask for the
probable values of the anomalous dimension or the condensate.

\bet
\begin{center}
\begin{tabular}{|c||c|c|c|c|c|}
\hline
 & $N=0$ & $N=2$ & $N=4$ & $N=8$ & $N=12$\\
\hline
$i=1$ & 0.642 & 2.002 &   3.320 & 5.903  &   8.480 \\
$i=2$ & 1.564 & 3.147 &   4.624 & 7.450  &   10.22 \\
$i=3$ & 2.537 & 4.241 &   5.800 & 8.769  &   11.64 \\
$i=4$ & 3.525 & 5.281 &   6.904 & 9.978  &   12.95 \\
\hline
\end{tabular}
\end{center}
\caption{The values of the first four $z_i$ for the non-chiral random matrix model 
for $N=0,2,4,8$ and $12$.}
\eet{rmta}

For $N < N^*$, we will be able to further substantiate our results
via a comparison with the appropriate random matrix models.  Random
matrix models appropriate for describing low-lying eigenvalues in
a three dimensional gauge theory coupled to massless fermions that
generates a non-zero bilinear condensate can be found
in~\cite{Verbaarschot:1994ip} where the fermion operator $C$ for
each fermion flavor is realized as a random
anti-Hermitian matrix. Under parity, $C\to C^\dagger$ and therefore these random
matrix models will be parity-invariant for even number of flavors
since the Haar measure of a random Hermitian matrix is parity-invariant. 
Since our gauge group is $SU(2)$, $C$ is a real-symmetric $M \times M$ matrix in the random matrix model defined 
by~\cite{Magnea:1999iv,Nagao:2000um}
\be
Z = \int [dC] e^{-\frac{\pi^2}{16M} \Tr C^2} {\det}^{N} C,
\ee
and it is assumed that $M$ is taken to infinity.
Ordering of the eigenvalues of $C$ is according
to the absolute value of $C$ since this matches with the definition in a parity invariant
theory.  We opted to numerically simulate the random matrix model, and
the universal numbers, $z_i$, appearing in \eqn{sigdef}
are the averages of the eigenvalues so ordered. We have listed their values
in \tbn{rmta}. 

\section{Results}\label{sec:results}
\bef
\includegraphics[scale=0.88]{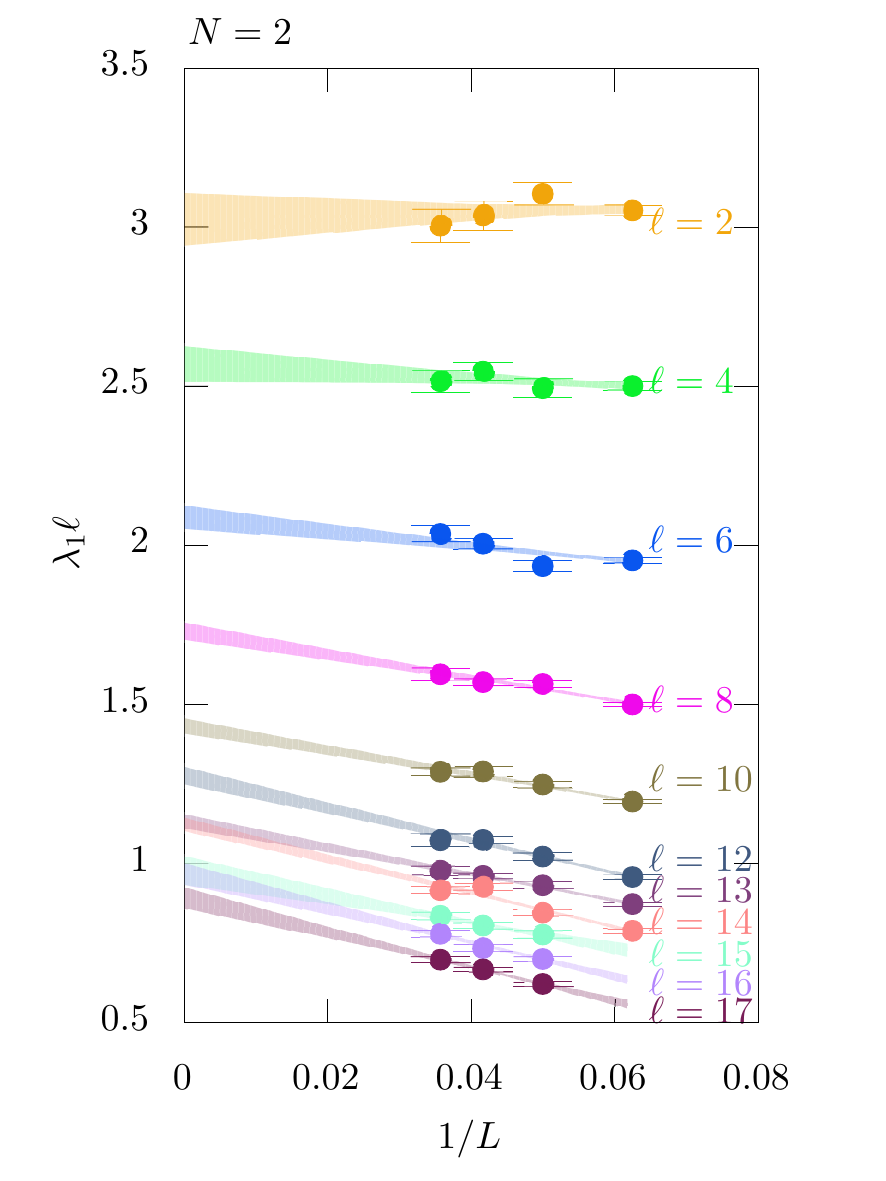}
\includegraphics[scale=0.88]{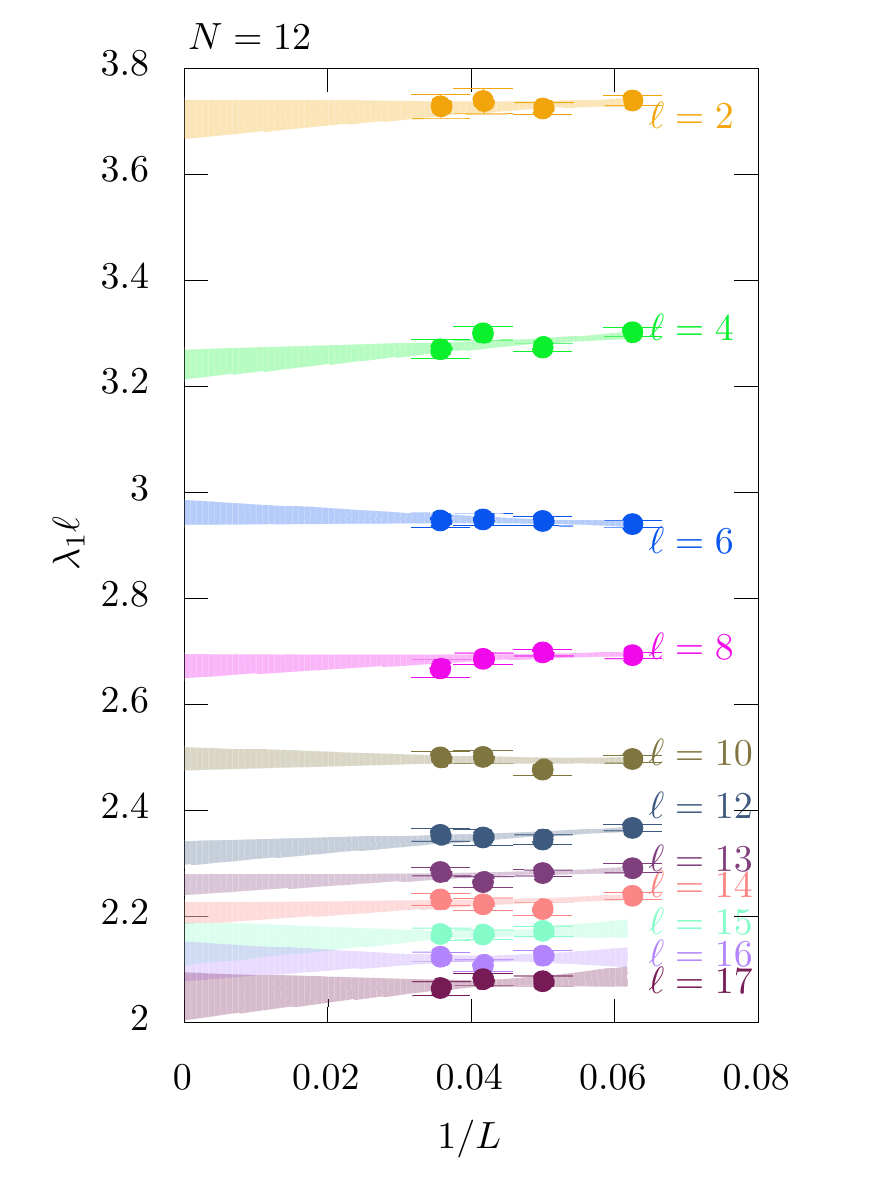}
\caption{
The continuum limits of $\lambda_1\ell$ at different fixed $\ell$
using a $1/L$ extrapolation for $N=2$ flavors is shown in the left panel and for
$N=12$ in the right panel.  The values of $\ell$ are given by the side of the data.
}
\eef{eigcont}
We present the results of our numerical analysis in this section.
We have given the list of our simulation points ($\ell$, $L$, $N$)
along with the corresponding averages of the first four eigenvalues
of the massless Hermitian Wilson-Dirac in Tables-\ref{tb:tab0},\ref{tb:tab2},\ref{tb:tab4},\ref{tb:tab8},\ref{tb:tab12} in~\apx{tabs}.
Using the eigenvalues at fixed $\ell$ at four different values of
$L$, we obtained the continuum eigenvalues, $\lambda_i(\ell)$;
$i=1,2,3,4$, by using a linear extrapolation in $1/L$. We have also
tabulated these continuum values in the same set of tables.
To illustrate the extrapolation pictorially, we show these continuum
extrapolations along with the 1-$\sigma$ error bands
 for $\lambda_1\ell$ for $N=2$
and $N=12$  in \fgn{eigcont}.

\bef
\begin{center}
\includegraphics[scale=0.85]{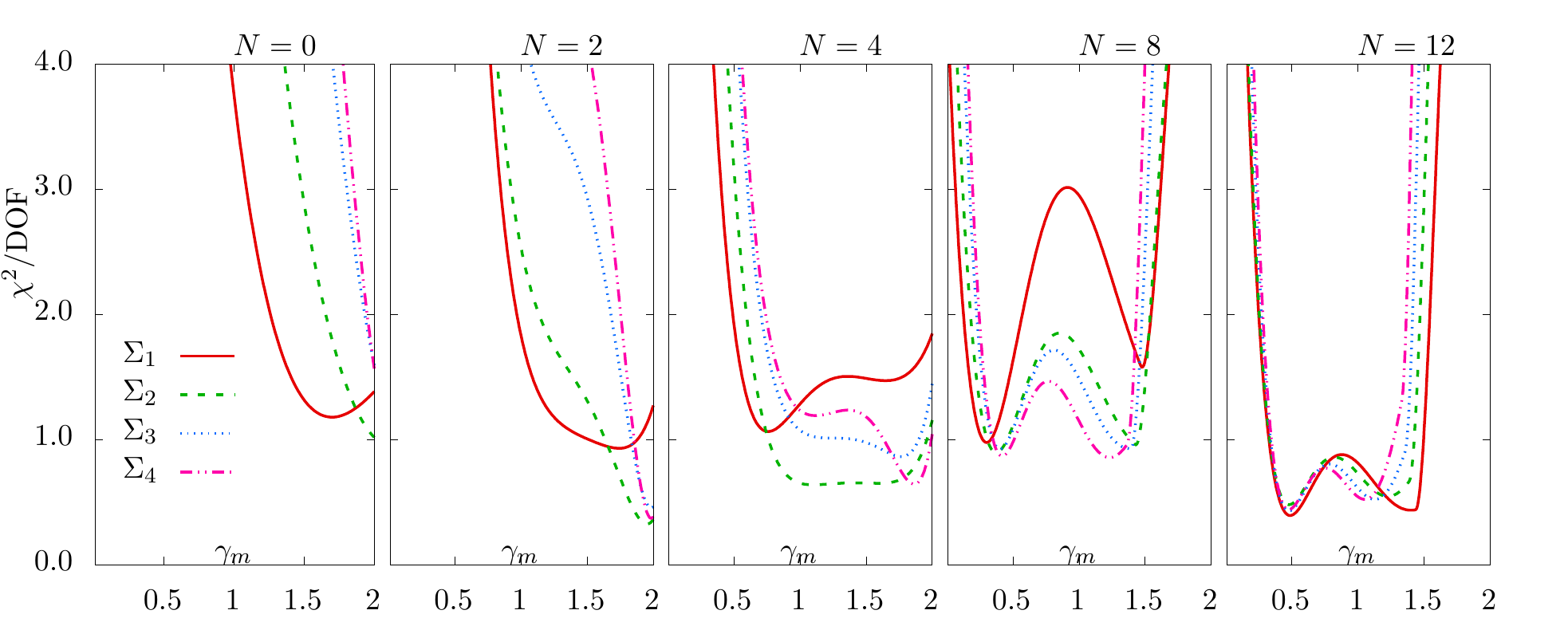}
\end{center}
\caption{A plot of $\chi^2$ per degree of freedom as a function of
$\gamma_m$ for different flavors $N$. The different curves are
obtained from various $\Sigma_i$.
}
\eef{chinf}

Using the continuum extrapolated values of $\lambda_i(\ell)$, we
obtained $\Sigma_i(\ell)$ as defined in \eqn{sigil}. We fit
$\Sigma_i(\ell)$ as a function of $\ell$ using \eqn{sigfit} with
different values of the exponent $\gamma_m$.  In \fgn{chinf}, we
show the $\chi^2/{\rm DOF}$ for these fits as a function of $\gamma_m$
for different $N$, as labelled on top of each panel.  The different curves in the panels 
correspond to $\chi^2/{\rm DOF}$ of the fits to the four different $\Sigma_i$. 
In order to easily interpret the plot, a rule
of thumb is that the fit using a value of $\gamma_m$ is good if its
$\chi^2/{\rm DOF}$ is about 1, while it being 2 or above is indicative
of the fit describing the data poorly. In the different panels, 
the limit $\gamma_m\to 2$ points to a theory
with a bilinear condensate and the limit $\gamma_m\to 0$
points to a free field behavior. 

For the $N=0$ and $N=2$ theories, the $\chi^2/{\rm DOF}$ has a
minimum around $\gamma_m=2$, thereby favoring a non-zero bilinear
condensate.  The $N=8$ and $N=12$ theories clearly disfavor
$\gamma_m=2$, instead favoring a scale-invariant behavior with a
non-trivial anomalous dimension.  For both these $N$, the $\chi^2$
minima are seen at two values of $\gamma_m$; one at $\gamma_m\approx
0.4$ and another at $\gamma_m\approx 1.4$. The two allowed values
of $\gamma_m$ separated by 1, points to the possibility that the
$\ell^{-2+0.4}$ behavior of $\Sigma_i(\ell)$ describing the data
at large values of $\ell$ could either correspond to the leading
term in \eqn{sigfit}, or it could be the subleading term in
\eqn{sigfit} which is dominant in the range of $\ell$ where we have
the data, with the leading $\ell^{-2+1.4}$ term becoming dominant
only at even larger $\ell$ where we do not have the data.  If one
assumes a well-behaved $1/\ell$ expansion with successively smaller
higher order terms, with no cross-over from one type of leading
behavior to another, one would favor the smaller of the allowed
values of $\gamma_m$, which is around 0.4 to 0.5 for both $N=8$ and
12.  At all flavors, the allowed range of $\gamma_m$ as deduced from
$\Sigma_1$ is broader than as allowed by other $\Sigma_i$.  The most likely cause is that
the behavior of the lowest eigenvalue is affected
the most by the need to fine tune the Wilson mass to realize massless
fermions on the lattice as explained in \apx{numdet}.  This is
enhanced at $N=4$ as is evident from the flat behavior of the
$\chi^2$ for $i=1$ at $N=4$.  For $N=4$, the range of allowed
$\gamma_m$ as deduced from the other $i$ are also broad and includes
$\gamma_m=2$, and hence we are unable to rule out scale-breaking in this case.
In the following subsections, we analyze the different $N$ separately.

\subsection{$N=0$}\label{sec:n0}
\bef
\includegraphics[scale=1.1]{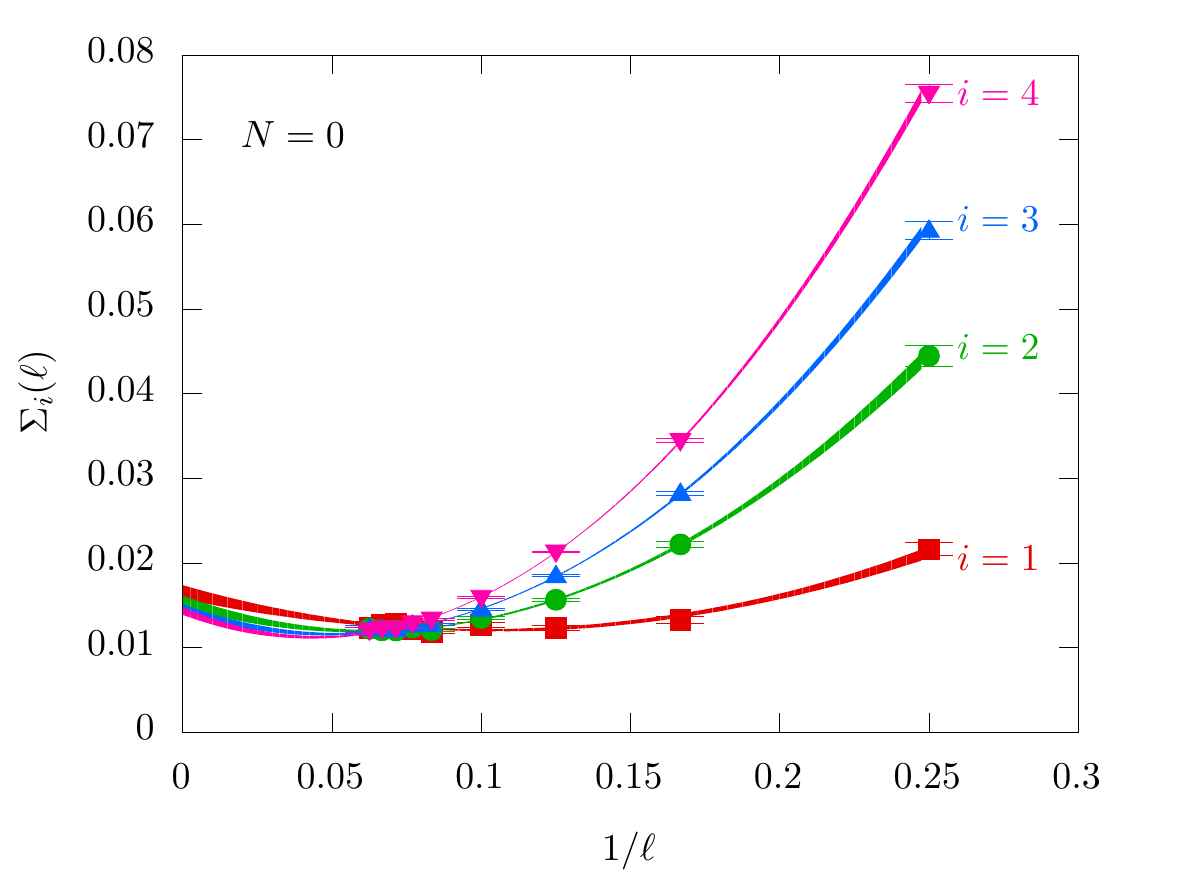}
\caption{ A plot of $\Sigma_i(\ell)$ as a function of $\ell$ for
$N=0$ flavors \ie, the quenched limit.  The different colors
correspond to different $i$ as specified in the plot. The data as
well as the error-bands of the fit using \eqn{sigcond} are shown.
The different $\Sigma_i$ extrapolate to about the same value in the
infinite volume limit.
}
\eef{signf0}

Using the four values of $z_i$ for $N=0$ in \tbn{rmta}, we obtained
$\Sigma_i(\ell)$ from \eqn{sigil}.  In \fgn{signf0}, we have plotted
$\Sigma_i(\ell)$ as a function of $1/\ell$.  One can see 
that the asymptotic behavior has set in for $\ell\ge 14$ with a finite
value of the bilinear condensate in the $\ell\to\infty$ limit. In
the same plot, we have also shown the error bands of fits to the
data using \eqn{sigcond}. We find the  values of $\Sigma_i$
extrapolated to infinite $\ell$ for $i=1,2,3,4$ to be $0.0165(8),
0.0154(7), 0.0147(5)$ and $0.0143(4)$ respectively.  It is reassuring
that the values of $\Sigma_i$ for different $i$ converge to about the
same value within errors in the $\ell\to\infty$ limit.  By taking
the average over all the four values of $\Sigma_i(\ell=\infty)$,
we quote the value of the bilinear condensate per color degree of
freedom for $N=0$ as
\be
\frac{\Sigma}{N_c} = 0.0076 \pm 0.0003;\quad N_c=2,
\ee
with the error being purely statistical.  We take the spread of
values in the four different $\Sigma_i$ to be a measure of the
systematic errors in the various extrapolations, and we conservatively 
estimate this systematic error to be about $0.0011$.  In the $N=0$ quenched
theory, the fermions are used merely as a probe with no back-reaction
on the gauge fields.  Since the pure $SU(2)$ theory does have a
scale, it is not a surprise to find a bilinear condensate in this
case.  We remind the reader that the above value of condensate
is dimensionless and in units of the coupling $g^2_{\rm ph}$.  When
the above value of condensate per color is measured in units of the
't Hooft coupling, $N_c g^2_{\rm ph}$, it is roughly a factor of
$2$ smaller than the corresponding value in the 
't Hooft limit in~\cite{Karthik:2016bmf}.
It is also interesting to note that a linear extrapolation  in
$\frac{1}{N_c}$ of the $N_c=1$ value in~\cite{Karthik:2017hol} and
the $N_c=2$ value obtained here of the quenched condensates in units
of 't Hooft coupling is consistent with the value in the 't Hooft
limit.

\subsection{$N=2$}\label{sec:n2}
\bef
\includegraphics[scale=1.1]{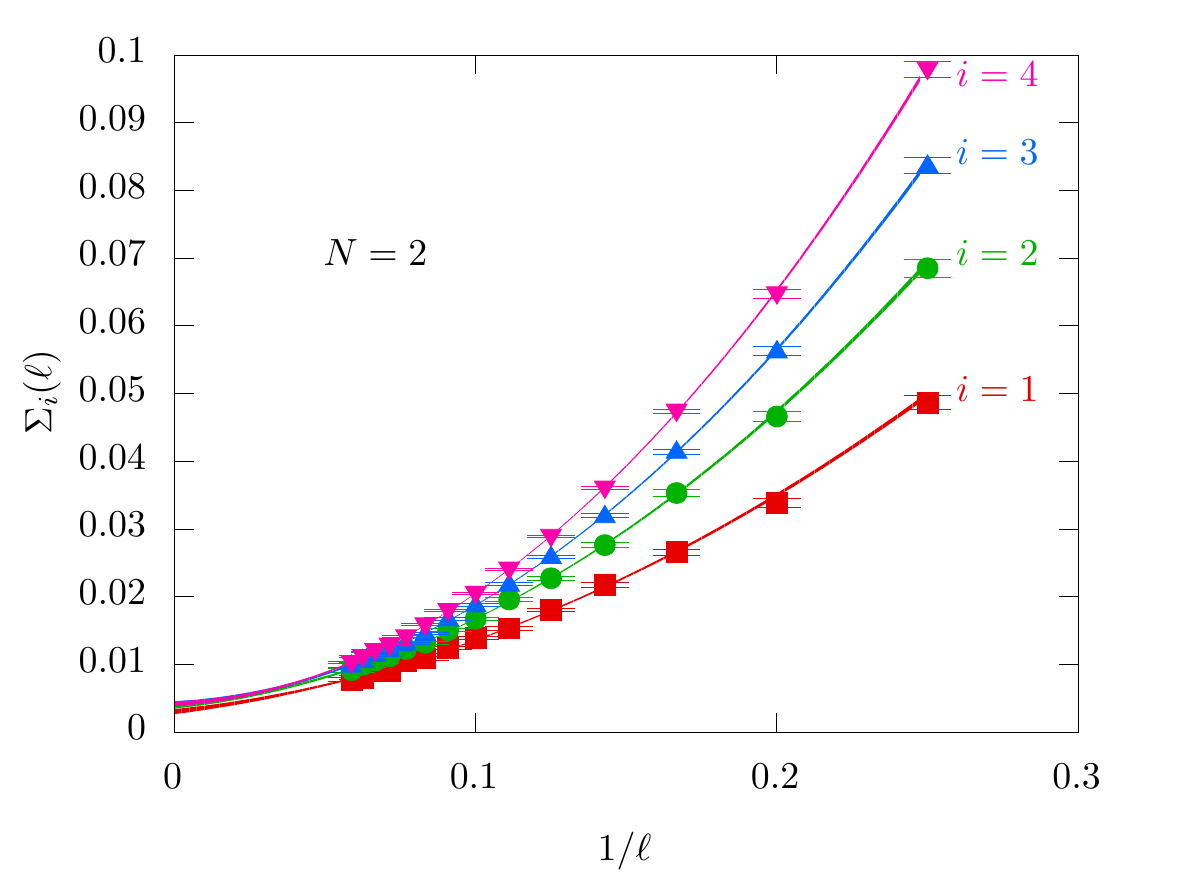}
\caption{ A plot of $\Sigma_i(\ell)$ as a function of $\ell$ 
for $N=2$, with the different symbols and error bands as explained in \fgn{signf0}.}
\eef{signf1}

We  use the four values of $z_i$ for $N=2$ in \tbn{rmta} to obtain
$\Sigma_i(\ell)$ from \eqn{sigil}.  In \fgn{signf1}, we show the
$\ell$ dependence of $\Sigma_i$.  Unlike the $N=0$ theory, we do
not see an asymptotic plateauing of the condensate even at the
largest $\ell$ we were able to simulate.  We have shown the fits
of the data to \eqn{sigcond} by the solid lines in the same plot
for the different values of $i$. From these extrapolations, we
estimate the condensate at infinite volume for  $i=1,2,3,4$ to be
0.0030(4), 0.0039(4), 0.0042(4), 0.0042(4) respectively.  We find
the condensates estimated from different eigenvalues to converge
to significantly non-zero values at $\ell=\infty$.  As in the case
of $N=0$, these extrapolated values of $\Sigma_i$ from $i=2,3$ and
4 agree within errors. But, the extrapolated central value of $\Sigma_1$ is
30\% lower than the others, but still significantly larger than
zero. We think this is due to the difficulty in tuning the Wilson
mass to obtain exactly massless fermions. One can rectify this in
a future study with overlap fermions. Taking the average over the
estimates of $\Sigma$ from all the four low-lying eigenvalues, the
condensate per color degree of freedom for $N=2$ is
\be
\frac{\Sigma}{N_c} = 0.0019 \pm 0.0002;\quad N_c=2,
\ee
with the error being purely statistical. Taking the spread in the
estimated values of $\Sigma_i/N_c$, a conservative estimate of the
systematic error is 0.0006.  Note that the value condensate mesured
in units of 't Hooft coupling is lower by a factor of $10$ compared
to the one in the 't Hooft limit.

\subsection{$N=4$}\label{sec:n4}

\bef
\includegraphics[scale=0.76]{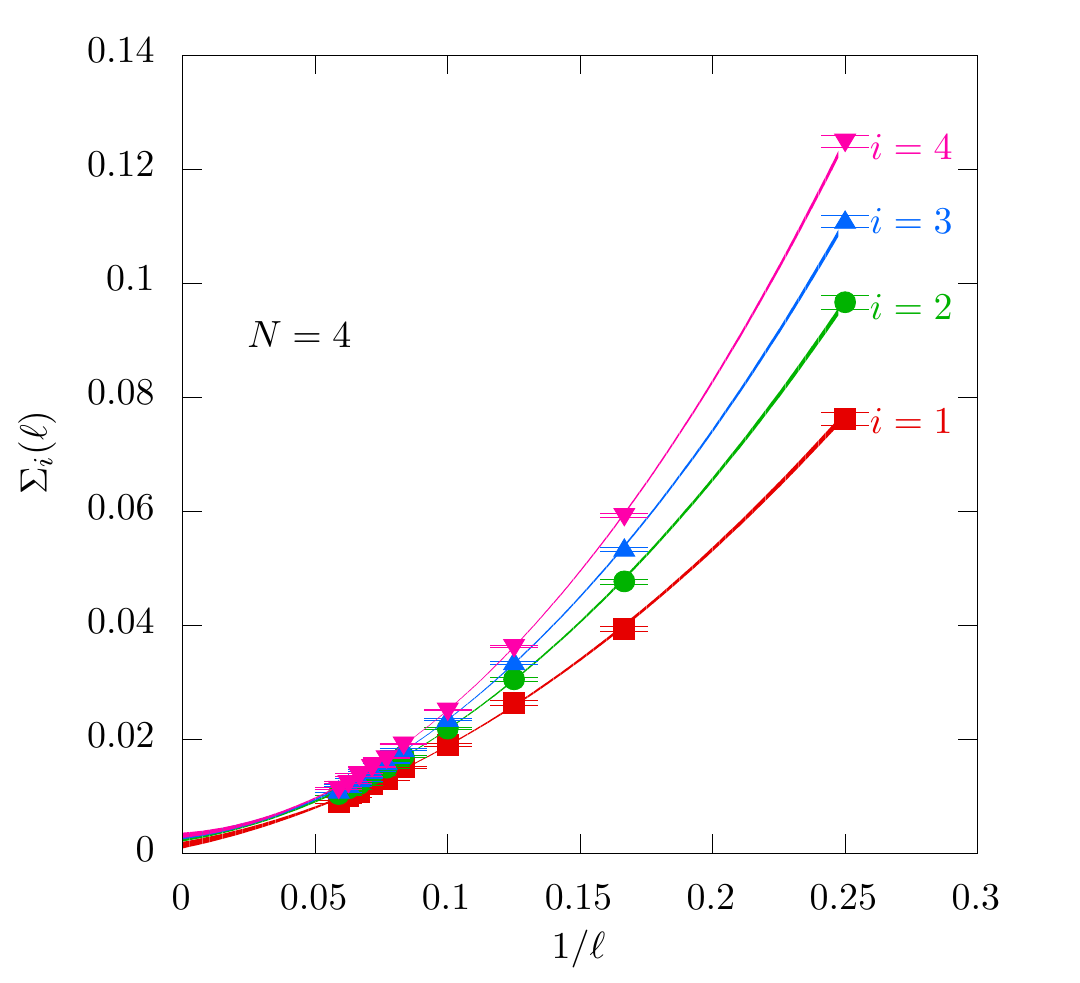}
\includegraphics[scale=0.75]{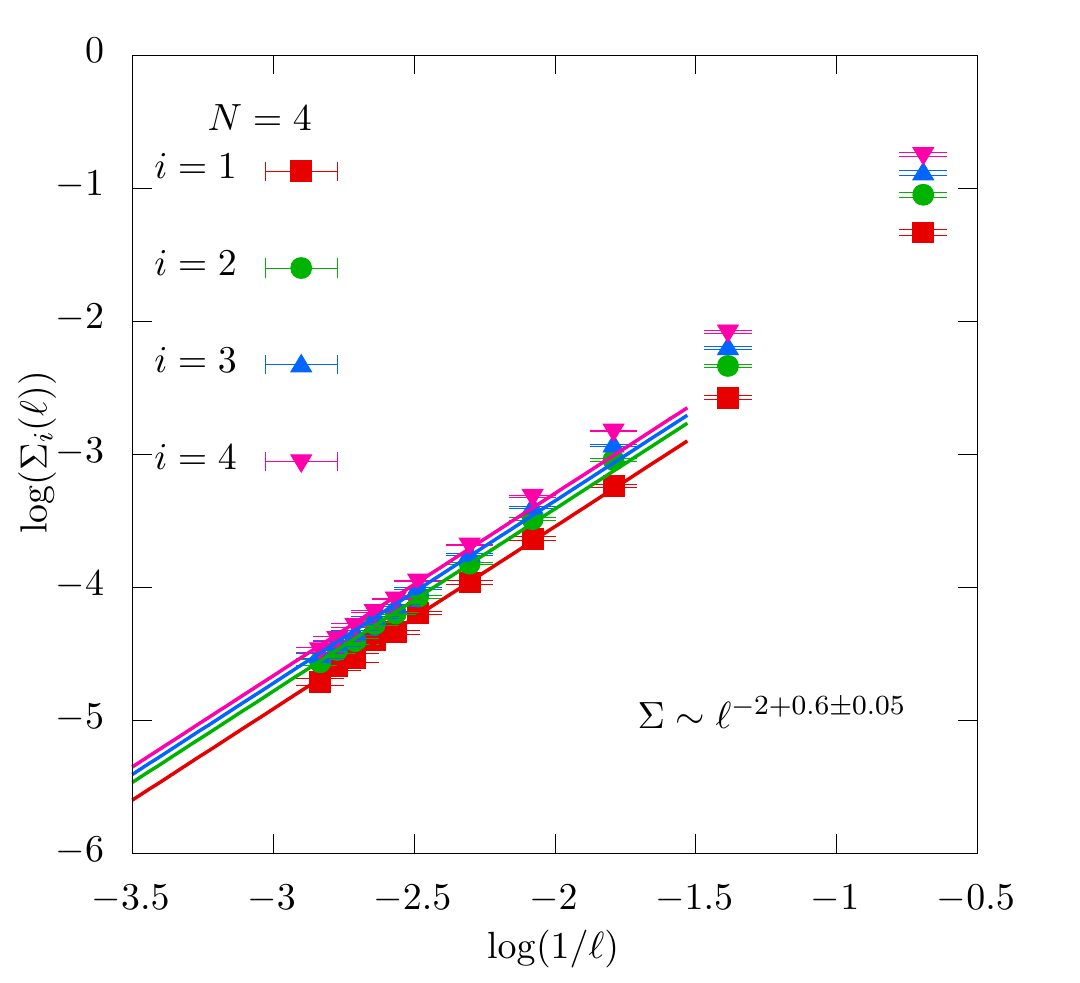}
\caption{ Plots of $\Sigma_i(\ell)$ as a function of $\ell$
 for $N=4$ in linear scale (left panel) and in log-log scale (right
 panel).  The analysis on the left panel assumes the presence of a
 non-zero condensate in the infinite volume limit, and hence
 explicitly sets $\gamma_m=2$. The error bands for the infinite
 volume extrapolations are shown along with the data in the left
 panel.  The analysis on the right panel assumes the absence of a
 condensate, and hence uses a $\gamma_m\ne 2$.
A possible $\ell^{-2+\gamma_m}$ scaling behavior, with $\gamma_m\approx
0.6$, is seen in the large $\ell$ we simulated.
}
\eef{signf4}

\bef
\includegraphics[scale=0.765]{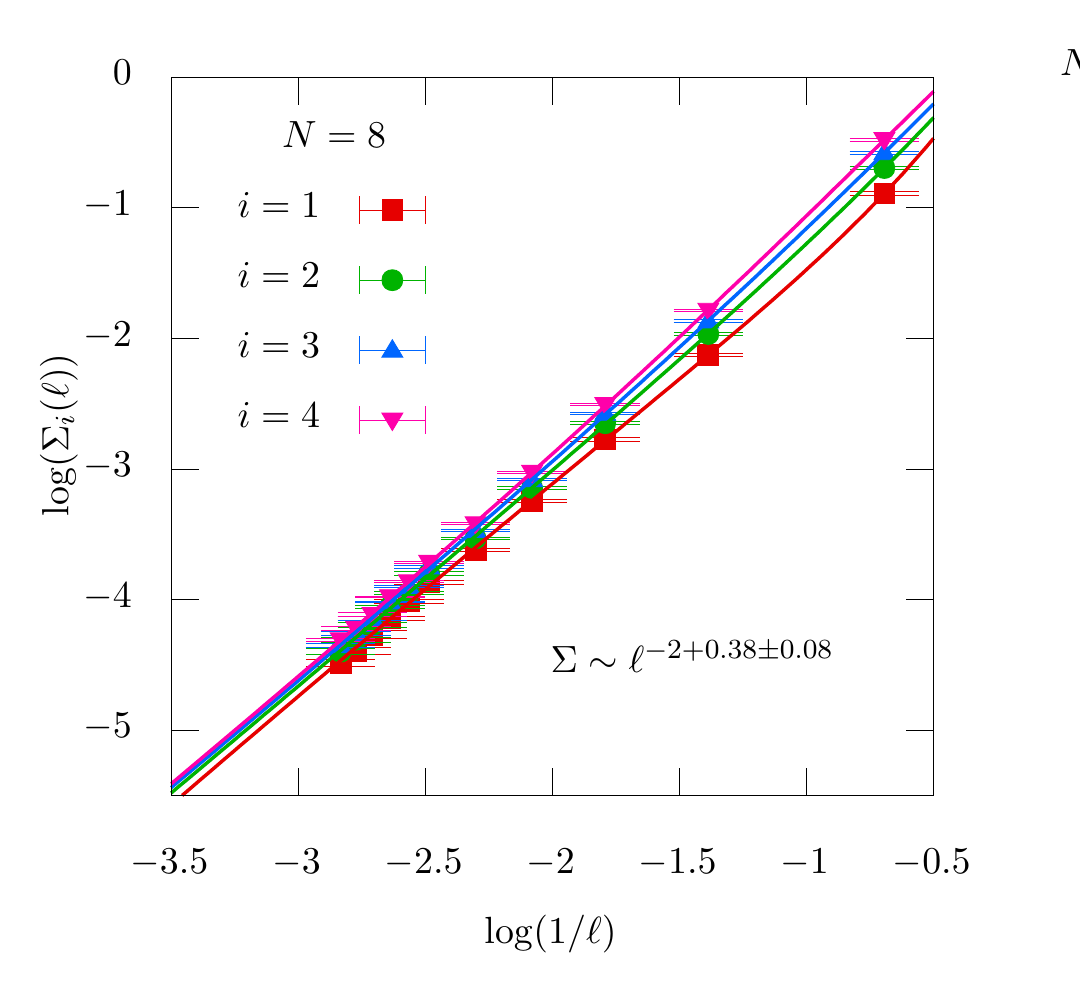}
\includegraphics[scale=0.765]{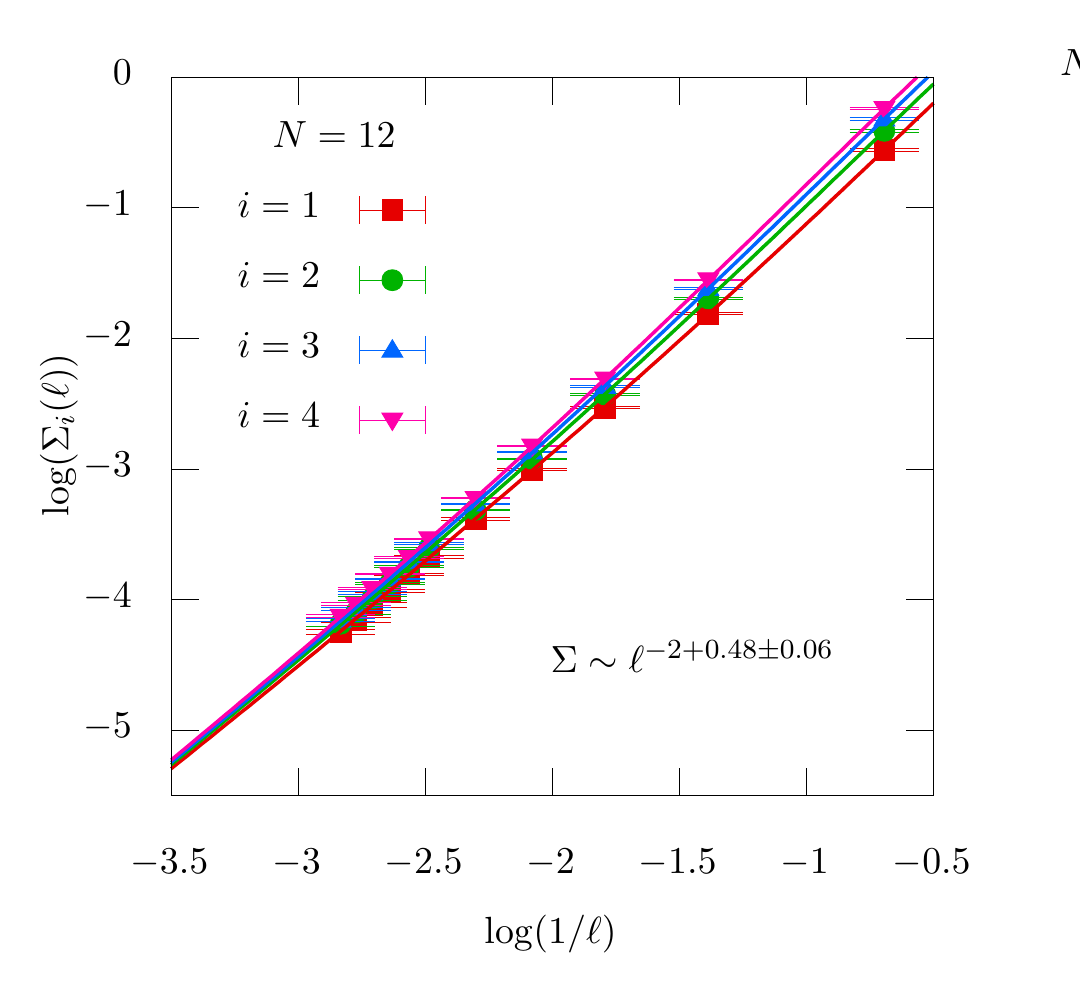}
\caption{ A log-log plot of $\Sigma_i(\ell)$ as a function of $\ell$
 for $N=8$ and $12$.  They clearly show a power-law behavior with
 the exponent $\gamma_m < 2$. This suggests $N=8$ and 12 are
 scale-invariant theories.
}

\eef{signf4-12}

For $N=4$, we again use the four corresponding values
of $z_i$ in \tbn{rmta} to obtain $\Sigma_i(\ell)$ from \eqn{sigil}.
From the third panel of \fgn{chinf} which shows the $\chi^2/{\rm
DOF}$ as a function of $\gamma_m$ for $N=4$, we find a wide range
of values of $\gamma_m$, including $\gamma_m=2$, that fits the data
well.  Therefore, we analyze the data first assuming the presence
of non-vanishing condensate, in which case we force $\gamma_m=2$
and extrapolate the condensate to infinite volume using \eqn{sigcond}.
This procedure is shown on a linear scale on the left panel of
\fgn{signf4}. This leads to the values 0.0015(6), 0.0026(6),
0.0030(5), 0.0032(5) for $\Sigma_1,\Sigma_2,\Sigma_3$ and $\Sigma_4$
respectively in the infinite volume limit.  We stress that
this analysis involves a prior assumption that $\gamma_m=2$, and
with this assumption we find a possible non-zero condensate. However,
on the right panel of \fgn{signf4}, we again show $\Sigma_i(\ell)$
as a function of $1/\ell$, but on a log-log plot; a power-law
behavior will be seen as a straight line on this scale with the
slope being the scaling exponent. Now we can see a possible scaling
behavior setting in for $\ell > 13$.  Assuming a perfect
$\ell^{-2+\gamma_m}$ behavior for $\ell>13$, we find $\gamma_m\approx0.60$
if the theory is scale-invariant.  Thus, the two different analyses
leads to two different conclusions.  A more careful analysis, using
even larger values of $\ell$ than we have used, could place this
theory on either side of the critical value $N^*$.

\subsection{$N=8$ and $12$}\label{sec:n4-12}

Having found an indication of transition from scale-broken to
conformal phase at $N\approx 4$, we studied $N=8$ and 12 to see if
we find strong evidence for a scale-invariant behavior. In
\fgn{signf4-12}, we show the behavior of $\Sigma_i(\ell)$ as a
function of $\ell$ in log-log plots.  For both $N=8$ and 12 we find
an asymptotic power-law behavior setting in for $\ell > 4$.
Incorporating the $1/\ell$ and $1/\ell^2$ corrections in \eqn{sigfit},
we find the values for $\gamma_m$ to be 0.38(8) and 0.48(6)  for
$N=8$ and 12 corresponding to the first minima of the two seen in
the last two panels of \fgn{chinf}. The fits to the data with these
exponents are also shown along with the data in \fgn{signf4-12}.
Thus $N \ge 8$ clearly lie in the scale-invariant phase. Even though
there is strong evidence for the presence of $\gamma_m\ne 2$, we
nevertheless performed an analysis assuming the presence of non-zero
$\Sigma_i$ using \eqn{sigcond}. We find the extrapolated values of
the condensates, after averaging over the four different estimates
$\Sigma_i$, to be $0(2)\times 10^{-6}$ and $0(7)\times 10^{-6}$ for
$N=8$ and 12 respectively, and hence consistent with a scale-invariant 
behavior.

\subsection{Combined analysis of condensate}
\bef
\includegraphics[scale=0.8]{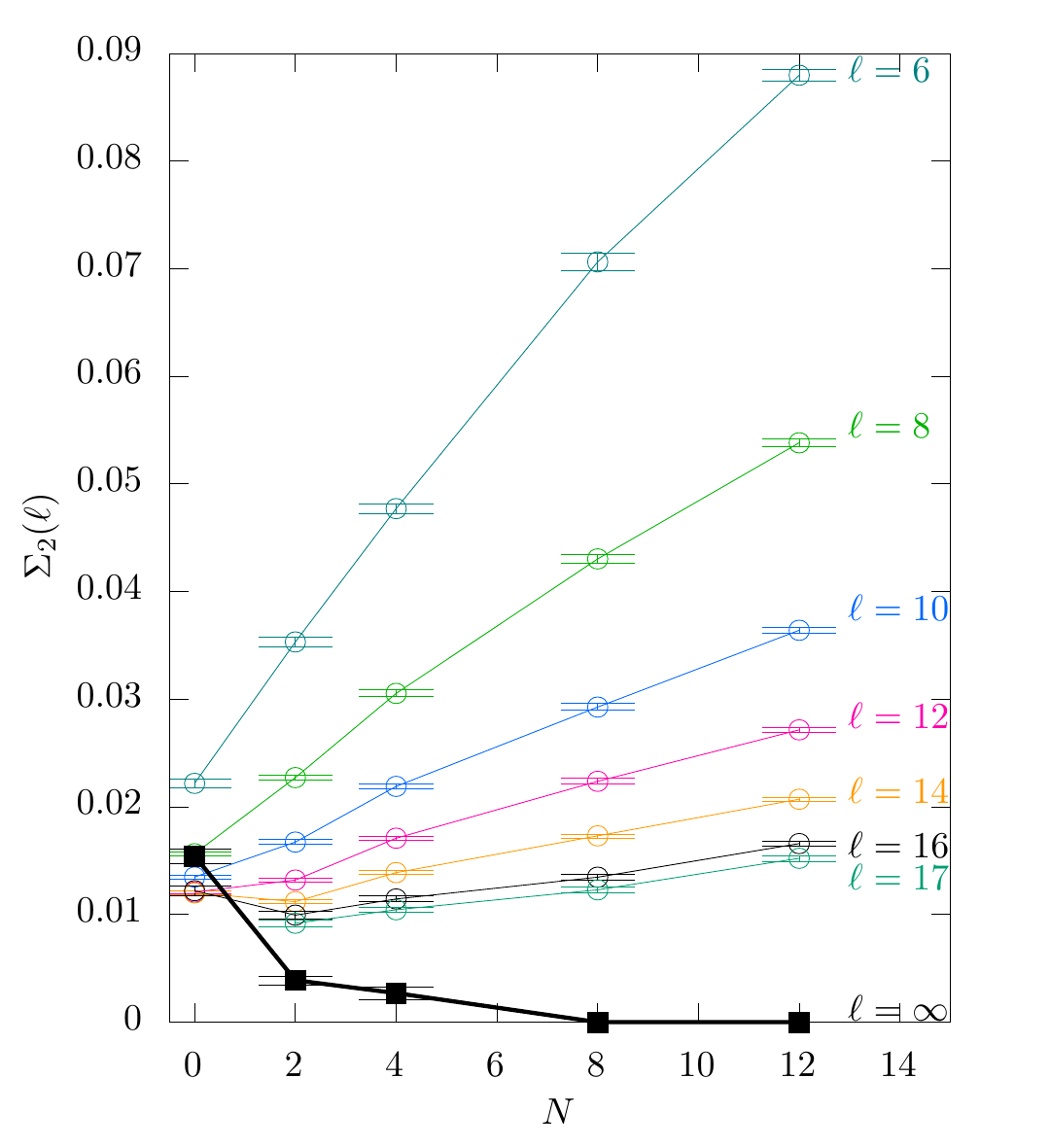}
\caption{ A plot of the condensate $\Sigma_2(\ell)$, as determined
from the second smallest eigenvalue, shown as a function of fermion
flavor $N$ at different fixed $\ell$, as labelled by the side.  The
open symbols correspond to the value
 of condensate determined at finite $\ell$, as defined in \eqn{sigil}.
 The black solid symbol corresponds to the extrapolated value of
 condensate at infinite $\ell$. The points are connected by lines
 to aid the eye.
}
\eef{sig2}

We consolidate our results on the condensate from different flavors
in \fgn{sig2}. For the sake of clarity, we have only shown the data
for the second eigenvalue.  The plot shows the condensate at finite
$\ell$ as we have defined using \eqn{sigil}. In the same plot, we
also show the expected value of condensate at infinite volume, with
the starting assumption that there is no non-trivial scaling dimension
present, which forces $\gamma_m=2$. As we have explained, this
is a good assumption for $N=0$ and 2, and we find a non-zero
condensate in these cases. While this is a bad assumption for $N=8$
and 12, the extrapolated values of the condensate in these cases
is nevertheless consistent with zero. However, our results are
inconclusive about the $N=4$ case; analyses assuming $\gamma_m=2$
as well as $\gamma_m < 2$ are consistent with the data. Thus, we
have shown a non-zero value of condensate in \fgn{sig2}. We could
not study any critical $\Sigma(N)\sim |N-N^*|^\Delta$ behavior near
$N^*\approx 4$ given the access to only two scale-broken integer
number of dynamical flavors.

\bef
\centering
\includegraphics[scale=0.85]{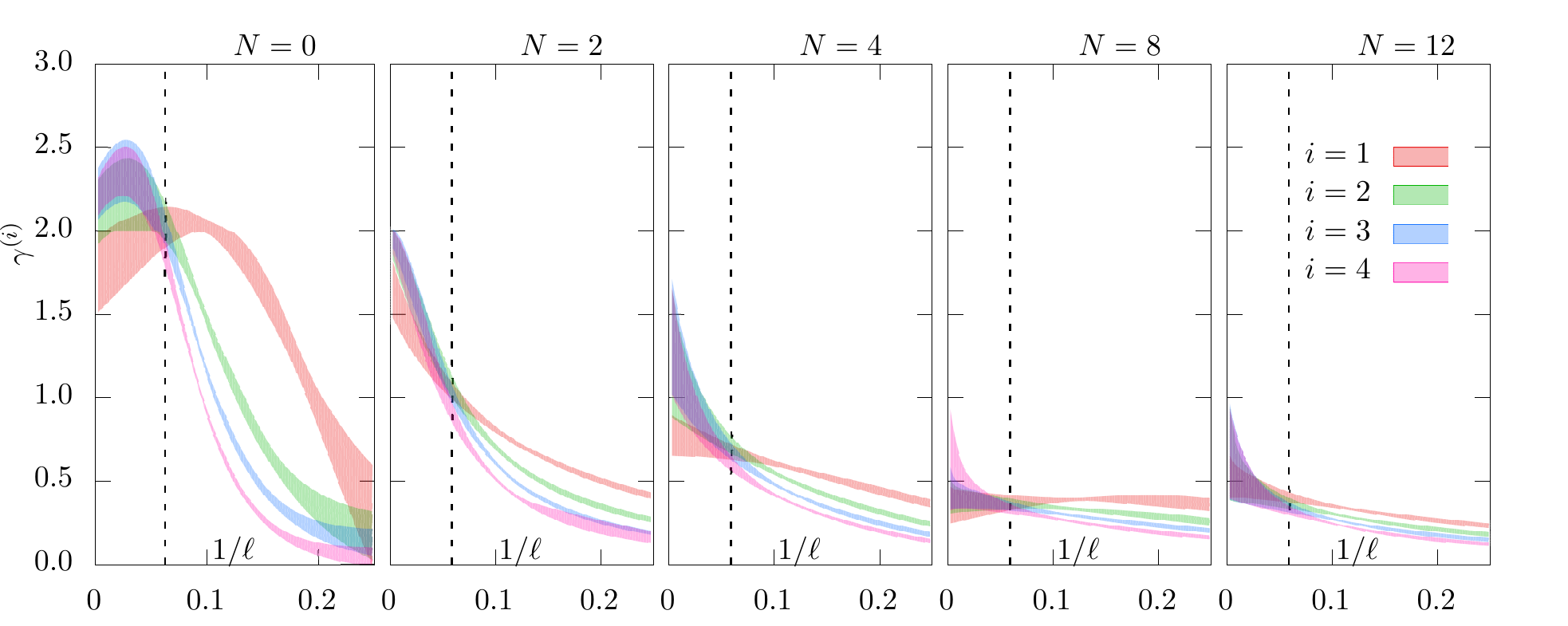}
\caption{The flow of $\gamma^{(i)}(\ell)$ as defined in \eqn{gammapdef} 
to its asymptotic value $\gamma_m$ in the infinite volume limit for $N=0,2,4,8$ and 12. 
The different colored error bands correspond $i=1,2,3,4$.
The dashed vertical lines correspond to the value of $1/\ell$ for the largest $\ell$ we have the data for, 
at each $N$.
}
\eef{flow}

\subsection{A flow from UV to IR}
As explained in the beginning of this section, we used \eqn{sigfit}
to describe the data and also to find the value of $\gamma_m$ that
best describes the asymptotic finite-size scaling behavior of the
low-lying eigenvalues. Once we have fit the data using \eqn{sigfit},
we can define an $\ell$ dependent $\gamma_m$ as
\be
\gamma^{(i)}(\ell)\equiv 2+\frac{\partial\log\left(\Sigma_i(\ell)\right)}{\partial \log(\ell)},
\label{gammapdef}
\ee
which will flow from $\gamma^{(i)}=0$ in the UV limit $\ell\to 0$,
to $\gamma^{(i)}=\gamma_m$ in the IR limit $\ell\to\infty$ for all
$i$.  In \fgn{flow}, we show this flow for all $N$.  As expected
the value of $\gamma^{(i)}(\ell)$ increases from values closer to
zero in smaller volumes as $\ell$ is increased. For $N=0$ and $N=2$,
in the infinite $\ell$ limit, the values of $\gamma^{(i)}$ converge
to values around 2, thereby showing the presence of condensate for
these smaller values of $N$. For $N=8$ and 12, $\gamma^{(i)}$ clearly
flows to asymptotic values smaller than 1, indicating the presence
of non-trivial mass anomalous dimension characteristic of infra-red
fixed points in these theories. For $N=4$, $\gamma^{(1)}$ seems to
flow to value less than 1, but $\gamma^{(i)}$ from larger $i$ flow
to values around 1.5 at $68\%$ confidence. However, at 94\% confidence
levels, the flow to $\gamma=2$ is also possible. So, this plot sums
up our restricted knowledge of the $N=4$ theory.  In order to give
confidence in the extrapolations to $\ell\to\infty$, in the different
panels, we have separated the range of $\ell$ where we have the
data from the range of larger $\ell$ where we do not.  The flow to
infinite $\ell$ which is extrapolative, is smooth and well-behaved.
In all the panels, in the range of $\ell$ where we have the data,
the values of $\gamma^{(i)}(\ell)$ from different $i$ do not in
general agree.  But $\gamma^{(i)}$ for $N=0,2,8$ and 12 converge
to values consistent with each other in the infinite $\ell$ limit
thereby giving confidence in the conclusions drawn in these cases.

\section{Conclusions}

We have performed a numerical analysis of three dimensional $SU(2)$
gauge theory coupled to an even number of massless fermions in such
a way that parity is preserved. We used Wilson fermions instead of
overlap fermions to reduce the computational cost.  The price to
pay was the absence of the full $Sp(N)$ flavor symmetry away from the continuum
limit but this did not prevent us from extracting the critical
number of fermion flavors.

We studied the finite volume behavior of the low lying Dirac spectrum
in a periodic finite physical volume of size $\ell^3$ using two
different forms, namely, \eqn{sigfit} and \eqn{sigcond}. The first
one has the anomalous dimension, $\gamma_m$, as a parameter that
we varied to find the best fit.  We found that the data clearly
favored $\gamma_m=2$ for $N=2$ and $N=4$ and the data clearly favored
$\gamma_m < 2$ for $N=8$ and $N=12$. Given the numerical limitations
in obtaining results at arbitrary large values of $\ell$, our data
showed it did not favor $\gamma_m=2$ for $N=4$. The presence of a
non-zero bilinear condensate for $N=0$ and $N=2$, and the absence
of one for $N=8$ and $N=12$ remained true when the data was analyzed
using \eqn{sigcond} which did not have the freedom of choosing a
$\gamma_m$ away from integer values. Analysis of the $N=4$ data
using \eqn{sigcond} suggests a small value for the bilinear condensate.

We therefore conclude that the critical number of flavors is somewhere
between $N=4$ or $N=6$ and we are not able to exclude $N=4$ using
the analysis presented in this paper. Having narrowed down the
critical value of the number of flavors to one of two integers, the
next step is to study the $N=4$ and $N=6$ theories using
overlap~\cite{Karthik:2016ppr} or domain-wall
fermions~\cite{Hands:2015dyp,Hands:2015qha}.  Since the $U(N)$
flavor symmetry as well as the larger $Sp(N)$ global symmetry 
will be exact in the lattice theory, the behavior
of all low lying eigenvalues can be used in the numerical analysis
with equal confidence. Furthermore, one can also study the propagator
of scalar and vector mesons and extract the behavior of their masses
in finite volume.

The analysis presented in this paper clearly shows that one can use
$SU(2)$ gauge theories with massless fermions in three dimensions
to numerically study the transition from scale invariant behavior
to one that generates a scale using continuum finite volume analysis.
It will also be interesting to test the predictions for symmetry breaking 
in~\cite{Komargodski:2017keh}

\acknowledgments
All computations in this paper were made on the JLAB computing clusters
under a class C project.  The authors acknowledge partial support
by the NSF under grant number PHY-1515446. 
N.K. also acknowledges partial 
support by the U.S. Department of Energy under contract No. DE-SC0012704.

\appendix

\section{Wilson fermions in three dimensions}\label{sec:numdet}
In our lattice simulation, both the gauge field as well as the $N$
flavors of fermions are dynamical. That is, the Boltzmann weight
$e^{-S}$ in our simulation using the Hybrid Monte Carlo (HMC) technique~\cite{Duane:1987de}
uses $S=S_f+S_g$, where $S_f$ and $S_g$ are the fermion and gauge
action respectively. The lattice variables are the SU$(2)$ gauge-links;
a gauge link $U_\mu(n)$ is an $SU(2)$ matrix that represents the
parallel transporter from site $n=(n_1,n_2,n_3)$ to site $(n+\hat
\mu)$, with $1\le n_i \le L$.  We impose periodic boundary condition
for both the gauge field as well as the fermions in all three
directions of the lattice; for SU$(2)$ theory this boundary condition
is sufficient since both $1$ as well as $-1$ are part of the gauge
group.  We use the standard single plaquette gauge action, namely,
\be
S_g = -\frac{2L}{\ell} \sum_n \sum^{3}_{\mu>\nu=1} \Tr \ P_{\mu\nu}(n),
\ee
where $P_{\mu\nu}(n)$ is the parallel transporter around a plaquette in $\mu\nu$-plane 
at lattice site $n$:
\be
P_{\mu\nu}=U_\mu(n)U_\nu(n+\hat\mu)U^\dagger_\mu(n+\hat\nu)U^\dagger_\nu(n).
\ee
 We reduced lattice spacing effect in 
fermion action as well as fermionic observables by smoothening the 
gauge field that enters the Dirac operator using the technique of gauge-link smearing.  
For this, 
we used 1-level improved stout links~\cite{Morningstar:2003gk}, denoted
by $V_\mu(n)$:
\be
V_\mu(n) = e^{i\frac{s}{4} Q_\mu(n)} U_\mu(n)\quad{\rm where}\quad
Q_\mu(n) =\sum_{\nu\ne \mu}\frac{1}{2i} \left(P_{\mu\nu}(n)-\frac{1}{2}\Tr P_{\mu\nu}(n)\right) - {\rm h.c}.
\ee
We used an optimum value $s=0.65$ where the value of the smeared plaquette was maximized.
For the regularized two-component Dirac operator in \eqn{hermd}, 
we used the two-component Wilson-Dirac operator $C_W$ given by
\be
C_W(n,m)=(-3+M_P)\delta_{n,m}+\frac{1}{2}\sum_{\mu=1}^3\left\{\left(1+\sigma_\mu\right)V_\mu(n)\delta_{n+\hat\mu,m}+\left(1-\sigma_\mu\right)V^\dagger_\mu(n-\hat \mu)\delta_{n-\hat\mu,m}\right\},
\ee
where $\sigma_\mu$ are the Pauli matrices, and $M_P$ is the Wilson
mass which needs to be fine-tuned to non-zero values at finite
lattice spacings in order study massless continuum fermions. 
The above two-component Wilson-Dirac operator satisfies 
$C^t_W(n,m)=\sigma_2\tau_2 C_W(n,m) \sigma_2\tau_2$, and hence there is a 
$Sp(1)$ symmetry associated with a single flavor of 
two-component Wilson fermion even at finite 
lattice spacing.
We
incorporated the resulting fermion determinant, $\det\left(C_W
C^\dagger_W\right)^{N/2}$, from the $N/2$ parity-invariant pairs
of two-component Wilson-Dirac fermions in our HMC simulation using
using $N/2$ pseudo-fermion random vectors as explained
in~\cite{Karthik:2015sgq}.  In the HMC, we were able to marginally
optimize the molecular dynamics stepsize by using the Omelyan
symplectic integrator~\cite{PhysRevE.65.056706} as well as by tuning
the stepsize at run-time such that the acceptance is above 80\%.

\bet
\centering
{\renewcommand{\arraystretch}{0.5}
\begin{tabular}{|c|c|c|c|c|c|}
\hline
$N$ & $L$ & $M_1\times 10^{2}$ & $M_2\times 10^3$ & $M_3\times 10^4$ \\
\hline
0  &  16 & 2.45 & 2.70 & 7.58 \\
   &  20 & 2.14 & 1.48 & 5.52 \\
   &  24 & 1.26 & 2.58 & 3.34 \\
   &  28 & 0.82 & 2.82 & 2.25 \\
   &  32 & 0.41 & 3.18 & 1.39 \\
\hline
2  &  16 & 3.09 & 3.66 & 4.80 \\
   &  20 & 1.82 & 3.62 & 3.14 \\
   &  24 & 1.25 & 3.60 & 1.89 \\
   &  28 & 0.69 & 3.79 & 1.11 \\
\hline
4  &  16 & 2.98 & 5.43 & 2.61 \\
   &  20 & 1.98 & 4.11 & 3.04 \\
   &  24 & 1.21 & 4.19 & 1.14 \\
   &  28 & 0.96 & 3.62 & 0.84 \\
\hline
8 &   16 & 3.33 & 5.92 & 1.08 \\
  &   20 & 2.08 & 4.90 & 0.82 \\
  &   24 & 1.41 & 4.21 & 0.60 \\
  &   28 & 0.98 & 3.79 & 0.37 \\
\hline  
12 &  16 & 3.91 & 4.78 & 1.18 \\
   &  20 & 2.36 & 4.41 & 0.66 \\
   &  24 & 1.72 & 3.72 & 0.49 \\
   &  28 & 1.12 & 3.57 & 0.27 \\
\hline
\end{tabular}
}
\caption{Table of values for $M_1,M_2$ and $M_3$ which parametrize the $\ell$ dependence of the tuned Wilson mass $M_P$ as given in \eqn{mpparam}. The values at different $L$ and $N$
are tabulated.}
\eet{mp}

As discussed earlier, the eigenvalues of the four-component Hermitian Wilson-Dirac operator, $D_W$,
appear in $\pm$ pairs. The operator $C_W$ is not anti-Hermitian but
becomes essentially one upon tuning $M_P$ to achieve massless
fermions. As such, the $Sp(1)\times Sp(1)$ symmetry at finite 
lattice spacing  becomes the full $Sp(2)$ symmetry only in the continuum limit.
But, the positive eigenvalues of $D_W$ can be
used to study the presence or absence of a bilinear condensate as
discussed in the context of 
$U(N)$ flavor symmetry~\cite{Karthik:2015sgq}.  In order to realize massless
fermions, we tuned the value of $M_P$ at each simulation point to
that value where the lowest eigenvalue of $C_WC^\dagger_W$ is
minimized when measured over a small ensemble of thermalized
configurations at that simulation point.  Since these tuned $M_P$
are required for any future computation at larger $\ell$ and $L$,
and also for normalizing the eigenvalues of the overlap operator
which makes use of the Wilson-Dirac kernel~\cite{Karthik:2016ppr}, we
parametrize the tuned value of
$M_P(\ell)$ for different $L$ and $N$ using 
\be
M_P(\ell)= M_1+M_2\ell+M_3\ell^2,
\label{mpparam}
\ee
and tabulate these parameters in~\tbn{mp}.

We used Ritz algorithm~\cite{Kalkreuter:1995mm} to compute the four
low-lying eigenvalues of $\sqrt{C^\dagger_W C_W}$.  We measured the
eigenvalues every five trajectories of HMC after thermalization,
and this way we collected about 3500 to 4500 measurements at different
$\ell$, $L$ and $N$. We accounted for autocorrelations in the data
by using blocked jack-knife error analysis. The simulation points at
$N=0,2,4,8$ and 12 along with the low-lying eigenvalue measurements
are tabulated in \apx{tabs}. 

\section{Tables of measurements}
\label{sec:tabs}
In the following tables, we have given the values and the errors
of the four low-lying eigenvalues of the Dirac operator for $N=0,2,4,8$
and 16. For each physical $\ell$, we have given these measurements
made using different $L^3$ lattices.  The $L=\infty$ values are the
continuum values obtained through a linear $1/L$ extrapolation.

\bet[H]
\centering
{\renewcommand{\arraystretch}{0.5}
\begin{tabular}{|c|l|l|l|l|l|}
\hline
$\ell$ & $L$ & $\lambda_1\ell$ & $\lambda_2\ell$ & $\lambda_3\ell$ & $\lambda_4\ell$ \\
\hline
4 &16& 1.942(29) & 2.295(27) & 2.748(20) & 3.004(18)\\
 &20& 1.892(43) & 2.246(42) & 2.711(28) & 2.965(27)\\
 &24& 1.902(46) & 2.249(44) & 2.723(33) & 2.971(31)\\
 &28& 1.917(37) & 2.261(34) & 2.722(26) & 2.973(24)\\
 &32& 1.891(56) & 2.246(53) & 2.704(41) & 2.959(37)\\
 &$\infty$& 1.857(66) & 2.199(61) & 2.674(46) & 2.919(42)\\
6 &16& 1.227(18) & 1.902(16) & 2.454(11) & 2.816(9)\\
 &20& 1.267(18) & 1.928(16) & 2.473(10) & 2.830(8)\\
 &24& 1.291(24) & 1.938(19) & 2.485(11) & 2.834(8)\\
 &28& 1.230(28) & 1.881(22) & 2.447(12) & 2.800(11)\\
 &32& 1.314(38) & 1.973(30) & 2.506(19) & 2.854(15)\\
 &$\infty$& 1.343(43) & 1.958(35) & 2.497(22) & 2.842(19)\\
8 &16& 0.693(9) & 1.419(9) & 2.003(6) & 2.440(5)\\
 &20& 0.727(10) & 1.473(7) & 2.044(7) & 2.480(6)\\
 &24& 0.736(15) & 1.467(18) & 2.059(12) & 2.492(9)\\
 &28& 0.759(12) & 1.485(12) & 2.065(11) & 2.505(8)\\
 &32& 0.727(17) & 1.473(16) & 2.065(10) & 2.509(6)\\
 &$\infty$& 0.814(20) & 1.562(20) & 2.144(15) & 2.585(10)\\
10 &16& 0.450(5) & 0.997(8) & 1.508(7) & 1.949(6)\\
 &20& 0.457(6) & 1.044(9) & 1.567(8) & 2.009(8)\\
 &24& 0.476(7) & 1.065(8) & 1.602(7) & 2.047(6)\\
 &28& 0.467(8) & 1.061(12) & 1.611(9) & 2.058(8)\\
 &32& 0.480(8) & 1.066(14) & 1.604(13) & 2.069(10)\\
 &$\infty$& 0.508(12) & 1.160(17) & 1.745(15) & 2.210(12)\\
12 &16& 0.320(3) & 0.705(5) & 1.097(5) & 1.465(6)\\
 &20& 0.327(3) & 0.748(6) & 1.163(6) & 1.550(6)\\
 &24& 0.340(4) & 0.777(6) & 1.211(7) & 1.605(6)\\
 &28& 0.339(5) & 0.776(8) & 1.209(9) & 1.604(9)\\
 &32& 0.350(5) & 0.805(8) & 1.237(9) & 1.646(8)\\
 &$\infty$& 0.375(7) & 0.899(11) & 1.389(12) & 1.832(13)\\
\hline
\end{tabular}
}
{\renewcommand{\arraystretch}{0.5}
\begin{tabular}{|c|l|l|l|l|l|}
\hline
$\ell$ & $L$ & $\lambda_1\ell$ & $\lambda_2\ell$ & $\lambda_3\ell$ & $\lambda_4\ell$ \\
\hline
13 &16& 0.279(2) & 0.600(4) & 0.937(5) & 1.260(6)\\
 &20& 0.285(3) & 0.634(6) & 1.000(9) & 1.347(10)\\
 &24& 0.293(3) & 0.658(7) & 1.031(7) & 1.386(7)\\
 &28& 0.291(4) & 0.658(6) & 1.036(7) & 1.404(8)\\
 &32& 0.294(5) & 0.673(7) & 1.062(11) & 1.432(14)\\
 &$\infty$& 0.312(6) & 0.748(9) & 1.187(12) & 1.613(13)\\
14 &16& 0.251(3) & 0.521(3) & 0.802(4) & 1.082(5)\\
 &20& 0.258(3) & 0.553(4) & 0.854(5) & 1.155(5)\\
 &24& 0.257(4) & 0.578(6) & 0.902(7) & 1.224(8)\\
 &28& 0.250(3) & 0.577(5) & 0.911(7) & 1.237(7)\\
 &32& 0.253(5) & 0.593(11) & 0.938(12) & 1.270(12)\\
 &$\infty$& 0.256(6) & 0.665(9) & 1.069(11) & 1.460(12)\\
15 &20& 0.226(3) & 0.482(4) & 0.750(5) & 1.016(6)\\
 &24& 0.226(3) & 0.497(4) & 0.779(5) & 1.052(6)\\
 &28& 0.228(4) & 0.513(6) & 0.804(7) & 1.095(8)\\
 &32& 0.225(3) & 0.516(6) & 0.813(9) & 1.110(11)\\
 &$\infty$& 0.226(8) & 0.578(13) & 0.927(18) & 1.274(21)\\
16 &20& 0.228(2) & 0.453(3) & 0.675(4) & 0.900(4)\\
 &24& 0.203(3) & 0.441(4) & 0.687(4) & 0.931(4)\\
 &28& 0.199(2) & 0.440(3) & 0.700(5) & 0.958(5)\\
 &32& 0.204(3) & 0.462(5) & 0.721(8) & 0.988(10)\\
 &$\infty$& 0.203(12) & 0.500(19) & 0.808(27) & 1.139(29)\\
 & & & & & \\
 & & & & & \\
 & & & & & \\
 & & & & & \\
 & & & & & \\
 & & & & & \\
 & & & & & \\
 & & & & & \\
\hline
\end{tabular}
}
\caption{Measurements for $N=0$.}
\label{}
\eet{tab0}

\bet[H]
\centering
{\renewcommand{\arraystretch}{0.5}
\begin{tabular}{|c|l|l|l|l|l|}
\hline
$\ell$ & $L$ & $\lambda_1\ell$ & $\lambda_2\ell$ & $\lambda_3\ell$ & $\lambda_4\ell$ \\
\hline
2 &16& 3.053(15) & 3.216(15) & 3.435(14) & 3.568(14)\\
 &20& 3.105(37) & 3.269(36) & 3.483(32) & 3.616(31)\\
 &24& 3.037(46) & 3.204(45) & 3.429(39) & 3.564(38)\\
 &28& 3.005(52) & 3.165(51) & 3.402(44) & 3.535(43)\\
 &$\infty$& 3.026(80) & 3.191(79) & 3.429(70) & 3.564(68)\\
3 &16& 2.771(10) & 3.001(10) & 3.266(9) & 3.439(9)\\
 &20& 2.701(36) & 2.934(36) & 3.216(30) & 3.388(30)\\
 &24& 2.799(42) & 3.022(42) & 3.286(35) & 3.460(33)\\
 &28& 2.783(45) & 3.004(45) & 3.274(38) & 3.446(36)\\
 &$\infty$&2.761(70) & 2.975(71) & 3.259(59) & 3.434(56)\\
4 &16& 2.501(14) & 2.796(14) & 3.112(11) & 3.322(10)\\
 &20& 2.494(29) & 2.790(30) & 3.113(25) & 3.319(23)\\
 &24& 2.546(28) & 2.839(28) & 3.151(24) & 3.357(21)\\
 &28& 2.516(34) & 2.818(33) & 3.121(26) & 3.330(25)\\
 &$\infty$& 2.570(56) & 2.870(55) & 3.167(44) & 3.372(41)\\
5 &16& 2.209(10) & 2.589(9) & 2.945(8) & 3.191(7)\\
 &20& 2.252(20) & 2.610(19) & 2.954(15) & 3.204(13)\\
 &24& 2.248(26) & 2.620(24) & 2.968(17) & 3.214(18)\\
 &28& 2.278(30) & 2.645(31) & 2.979(25) & 3.224(22)\\
 &$\infty$& 2.364(46) & 2.700(45) & 3.014(36) & 3.263(32)\\
6 &16& 1.953(8) & 2.393(7) & 2.776(6) & 3.061(5)\\
 &20& 1.935(17) & 2.377(18) & 2.771(14) & 3.048(12)\\
 &24& 2.005(17) & 2.426(16) & 2.798(11) & 3.073(9)\\
 &28& 2.036(25) & 2.439(24) & 2.815(20) & 3.086(17)\\
 &$\infty$& 2.089(36) & 2.475(34) & 2.841(26) & 3.094(21)\\
7 &16& 1.710(9) & 2.193(8) & 2.595(5) & 2.904(5)\\
 &20& 1.730(15) & 2.200(14) & 2.609(12) & 2.910(8)\\
 &24& 1.773(16) & 2.242(17) & 2.639(13) & 2.942(11)\\
 &28& 1.786(27) & 2.259(23) & 2.642(18) & 2.939(16)\\
 &$\infty$ & 1.881(36) & 2.324(34) & 2.708(26) & 2.989(22)\\
8 &16& 1.500(7) & 1.998(6) & 2.410(5) & 2.738(5)\\
 &20& 1.564(11) & 2.041(10) & 2.448(9) & 2.762(7)\\
 &24& 1.570(12) & 2.051(12) & 2.457(11) & 2.779(8)\\
 &28& 1.595(20) & 2.061(19) & 2.466(14) & 2.781(12)\\
 &$\infty$&1.730(26) & 2.164(25) & 2.555(22) & 2.852(18)\\
9 &16& 1.337(7) & 1.835(6) & 2.240(4) & 2.578(4)\\
 &20& 1.391(11) & 1.856(13) & 2.270(9) & 2.610(7)\\
 &24& 1.419(19) & 1.872(17) & 2.278(15) & 2.607(13)\\
 &28& 1.463(19) & 1.914(19) & 2.318(15) & 2.635(11)\\
 &$\infty$&1.612(30) & 1.981(29) & 2.396(23) & 2.707(19)\\
\hline
\end{tabular}
}
{\renewcommand{\arraystretch}{0.5}
\begin{tabular}{|c|l|l|l|l|l|}
\hline
$\ell$ & $L$ & $\lambda_1\ell$ & $\lambda_2\ell$ & $\lambda_3\ell$ & $\lambda_4\ell$ \\
\hline
10 &16& 1.195(6) & 1.655(6) & 2.060(6) & 2.400(5)\\
 &20& 1.248(10) & 1.708(12) & 2.101(9) & 2.435(8)\\
 &24& 1.289(17) & 1.754(15) & 2.139(12) & 2.473(10)\\
 &28& 1.288(13) & 1.723(17) & 2.131(15) & 2.463(11)\\
 &$\infty$&1.433(23) & 1.880(27) & 2.258(23) & 2.574(18)\\
11 &16& 1.074(7) & 1.502(7) & 1.891(6) & 2.220(7)\\
 &20& 1.132(11) & 1.554(10) & 1.940(8) & 2.272(8)\\
 &24& 1.162(14) & 1.580(13) & 1.962(10) & 2.289(9)\\
 &28& 1.178(13) & 1.591(15) & 1.969(12) & 2.301(13)\\
 &$\infty$&1.328(24) & 1.727(25) & 2.092(20) & 2.425(21)\\
12 &16& 0.958(7) & 1.357(6) & 1.727(6) & 2.056(6)\\
 &20& 1.022(11) & 1.424(13) & 1.789(11) & 2.110(11)\\
 &24& 1.074(12) & 1.469(11) & 1.835(10) & 2.155(10)\\
 &28& 1.073(21) & 1.445(22) & 1.830(19) & 2.138(15)\\
 &$\infty$& 1.277(27) & 1.656(27) & 2.025(23) & 2.310(22)\\
13 &16& 0.872(7) & 1.237(6) & 1.582(5) & 1.891(4)\\
 &20& 0.932(11) & 1.301(12) & 1.657(10) & 1.967(8)\\
 &24& 0.961(9) & 1.334(11) & 1.693(12) & 1.996(10)\\
 &28& 0.977(14) & 1.349(15) & 1.708(15) & 2.023(14)\\
 &$\infty$& 1.133(22) & 1.517(23) & 1.905(23) & 2.216(20)\\
14 &16& 0.788(7) & 1.129(7) & 1.451(6) & 1.743(5)\\
 &20& 0.846(9) & 1.186(7) & 1.517(9) & 1.816(9)\\
 &24& 0.926(11) & 1.255(12) & 1.575(12) & 1.876(11)\\
 &28& 0.916(11) & 1.248(11) & 1.575(11) & 1.879(11)\\
 &$\infty$& 1.123(21) & 1.434(20) & 1.768(21) & 2.092(19)\\
15 &20& 0.777(11) & 1.095(13) & 1.408(12) & 1.698(10)\\
 &24& 0.805(10) & 1.115(11) & 1.428(11) & 1.711(10)\\
 &28& 0.835(12) & 1.161(11) & 1.473(9) & 1.768(7)\\
 &$\infty$& 0.979(45) & 1.320(48) & 1.636(42) & 1.945(34)\\
16 &20& 0.700(7) & 1.007(10) & 1.299(10) & 1.566(9)\\
 &24& 0.734(11) & 1.031(13) & 1.323(13) & 1.598(13)\\
 &28& 0.779(10) & 1.075(12) & 1.372(10) & 1.645(8)\\
 &$\infty$& 0.967(34) & 1.237(44) & 1.546(40) & 1.842(33)\\
17 &20& 0.621(7) & 0.899(9) & 1.173(10) & 1.426(9)\\
 &24& 0.667(7) & 0.948(8) & 1.222(8) & 1.477(8)\\
 &28& 0.698(10) & 0.980(11) & 1.262(10) & 1.526(9)\\
 &$\infty$&0.892(33) & 1.186(38) & 1.484(38) & 1.773(36)\\
 & & & & & \\
 & & & & & \\
 & & & & & \\
\hline
\end{tabular}
}
\caption{Measurements for $N=2$.}
\eet{tab2}

\bet[H]
\centering
{\renewcommand{\arraystretch}{0.5}
\begin{tabular}{|c|l|l|l|l|l|}
\hline
$\ell$ & $L$ & $\lambda_1\ell$ & $\lambda_2\ell$ & $\lambda_3\ell$ & $\lambda_4\ell$ \\
\hline
2 &16& 3.351(16) & 3.509(16) & 3.693(14) & 3.814(14)\\
 &20& 3.309(23) & 3.464(22) & 3.657(20) & 3.778(19)\\
 &24& 3.286(31) & 3.445(31) & 3.638(27) & 3.761(26)\\
 &28& 3.260(44) & 3.421(43) & 3.609(39) & 3.736(38)\\
 &$\infty$& 3.146(63) & 3.305(61) & 3.514(56) & 3.642(54)\\
4 &16& 2.778(12) & 3.045(11) & 3.313(9) & 3.503(9)\\
 &20& 2.768(19) & 3.037(18) & 3.302(17) & 3.492(15)\\
 &24& 2.741(18) & 3.010(17) & 3.283(14) & 3.475(13)\\
 &28& 2.776(24) & 3.046(25) & 3.317(19) & 3.500(19)\\
 &$\infty$& 2.722(39) & 2.990(38) & 3.270(31) & 3.455(30)\\
6 &16& 2.286(10) & 2.655(8) & 2.970(7) & 3.215(6)\\
 &20& 2.293(13) & 2.666(13) & 2.983(11) & 3.224(9)\\
 &24& 2.285(12) & 2.654(11) & 2.974(10) & 3.213(7)\\
 &28& 2.342(18) & 2.693(17) & 3.009(14) & 3.241(14)\\
 &$\infty$& 2.343(29) & 2.692(26) & 3.019(22) & 3.234(19)\\
8 &16& 1.880(8) & 2.303(7) & 2.655(6) & 2.934(4)\\
 &20& 1.899(12) & 2.325(10) & 2.671(8) & 2.952(6)\\
 &24& 1.903(12) & 2.313(10) & 2.660(9) & 2.935(8)\\
 &28& 1.923(19) & 2.344(18) & 2.693(14) & 2.957(12)\\
 &$\infty$& 1.964(28) & 2.365(25) & 2.706(20) & 2.972(16)\\
10 &16& 1.568(7) & 2.006(6) & 2.359(5) & 2.651(5)\\
 &20& 1.616(9) & 2.032(8) & 2.387(6) & 2.680(5)\\
 &24& 1.613(9) & 2.037(7) & 2.391(7) & 2.680(5)\\
 &28& 1.668(15) & 2.063(18) & 2.413(13) & 2.688(11)\\
 &$\infty$& 1.747(22) & 2.111(20) & 2.470(17) & 2.744(15)\\
12 &16& 1.362(8) & 1.770(8) & 2.115(6) & 2.404(5)\\
 &20& 1.407(12) & 1.805(14) & 2.145(9) & 2.426(11)\\
 &24& 1.416(9) & 1.808(8) & 2.143(8) & 2.428(7)\\
 &28& 1.433(11) & 1.815(10) & 2.157(10) & 2.449(9)\\
 &$\infty$& 1.527(21) & 1.879(20) & 2.210(17) & 2.494(16)\\
\hline
\end{tabular}
}
{\renewcommand{\arraystretch}{0.5}
\begin{tabular}{|c|l|l|l|l|l|}
\hline
$\ell$ & $L$ & $\lambda_1\ell$ & $\lambda_2\ell$ & $\lambda_3\ell$ & $\lambda_4\ell$ \\
\hline
13 &16& 1.272(5) & 1.661(5) & 1.992(5) & 2.274(5)\\
 &20& 1.322(7) & 1.691(6) & 2.017(6) & 2.306(4)\\
 &24& 1.346(12) & 1.713(11) & 2.049(9) & 2.327(7)\\
 &28& 1.368(15) & 1.730(13) & 2.063(11) & 2.341(10)\\
 &$\infty$& 1.502(22) & 1.818(20) & 2.152(17) & 2.432(16)\\
14 &16& 1.204(9) & 1.581(8) & 1.893(6) & 2.168(5)\\
 &20& 1.220(10) & 1.588(9) & 1.914(8) & 2.193(7)\\
 &24& 1.272(10) & 1.623(8) & 1.940(8) & 2.216(8)\\
 &28& 1.267(12) & 1.629(10) & 1.952(9) & 2.226(8)\\
 &$\infty$& 1.372(22) & 1.697(20) & 2.030(17) & 2.306(15)\\
15 &20& 1.175(7) & 1.526(6) & 1.835(5) & 2.109(6)\\
 &24& 1.196(12) & 1.547(12) & 1.863(10) & 2.130(10)\\
 &28& 1.239(15) & 1.573(13) & 1.876(13) & 2.141(11)\\
 &$\infty$& 1.369(45) & 1.683(41) & 1.987(36) & 2.226(35)\\
16 &20& 1.103(7) & 1.442(7) & 1.736(6) & 1.998(5)\\
 &24& 1.140(9) & 1.473(9) & 1.769(8) & 2.035(8)\\
 &28& 1.144(14) & 1.472(13) & 1.772(12) & 2.036(11)\\
 &$\infty$& 1.278(40) & 1.578(38) & 1.888(34) & 2.162(31)\\
17 &20& 1.052(7) & 1.382(8) & 1.667(8) & 1.919(8)\\
 &24& 1.064(10) & 1.376(9) & 1.676(9) & 1.931(9)\\
 &28& 1.123(9) & 1.436(10) & 1.721(10) & 1.973(11)\\
 &$\infty$& 1.276(34) & 1.533(35) & 1.831(36) & 2.085(35)\\
 & & & & & \\
 & & & & & \\
 & & & & & \\
 & & & & & \\
 & & & & & \\
 & & & & & \\
 & & & & & \\
 & & & & & \\
\hline
\end{tabular}
}
\caption{Measurements for $N=4$.}
\eet{tab4}

\bet[H]
\centering
{\renewcommand{\arraystretch}{0.5}
\begin{tabular}{|c|l|l|l|l|l|}
\hline
$\ell$ & $L$ & $\lambda_1\ell$ & $\lambda_2\ell$ & $\lambda_3\ell$ & $\lambda_4\ell$ \\
\hline
2 &16& 3.597(18) & 3.746(18) & 3.903(16) & 4.016(16)\\
 &20& 3.582(17) & 3.732(16) & 3.895(15) & 4.006(15)\\
 &24& 3.602(24) & 3.749(24) & 3.912(22) & 4.025(22)\\
 &28& 3.596(30) & 3.740(30) & 3.903(27) & 4.018(26)\\
 &$\infty$& 3.593(52) & 3.735(52) & 3.909(47) & 4.023(46)\\
4 &16& 3.084(13) & 3.323(13) & 3.541(11) & 3.706(10)\\
 &20& 3.098(13) & 3.330(13) & 3.545(10) & 3.710(10)\\
 &24& 3.092(18) & 3.332(16) & 3.549(15) & 3.715(13)\\
 &28& 3.084(18) & 3.317(18) & 3.537(16) & 3.705(14)\\
 &$\infty$& 3.097(35) & 3.326(34) & 3.545(30) & 3.715(27)\\
6 &16& 2.711(9) & 3.002(8) & 3.257(7) & 3.456(6)\\
 &20& 2.679(15) & 2.984(14) & 3.242(11) & 3.442(11)\\
 &24& 2.695(16) & 2.988(16) & 3.241(14) & 3.441(13)\\
 &28& 2.666(25) & 2.958(24) & 3.225(21) & 3.427(19)\\
 &$\infty$& 2.626(35) & 2.930(33) & 3.194(29) & 3.398(27)\\
8 &16& 2.379(10) & 2.711(10) & 2.986(9) & 3.207(7)\\
 &20& 2.382(10) & 2.715(8) & 2.992(8) & 3.215(6)\\
 &24& 2.374(13) & 2.708(12) & 2.989(9) & 3.213(8)\\
 &28& 2.375(15) & 2.708(14) & 2.983(10) & 3.206(10)\\
 &$\infty$& 2.370(27) & 2.705(27) & 2.985(21) & 3.215(19)\\
10 &16& 2.147(10) & 2.497(8) & 2.780(7) & 3.014(7)\\
 &20& 2.158(11) & 2.507(10) & 2.788(8) & 3.021(7)\\
 &24& 2.149(14) & 2.504(13) & 2.786(10) & 3.018(9)\\
 &28& 2.193(18) & 2.528(18) & 2.809(14) & 3.038(13)\\
 &$\infty$& 2.208(30) & 2.544(28) & 2.823(23) & 3.047(20)\\
12 &16& 1.957(10) & 2.301(8) & 2.589(7) & 2.821(6)\\
 &20& 1.962(12) & 2.318(12) & 2.596(10) & 2.834(8)\\
 &24& 1.946(14) & 2.289(13) & 2.584(11) & 2.825(9)\\
 &28& 1.972(19) & 2.316(18) & 2.592(15) & 2.833(13)\\
 &$\infty$& 1.963(30) & 2.312(27) & 2.590(23) & 2.845(20)\\
\hline
\end{tabular}
}
{\renewcommand{\arraystretch}{0.5}
\begin{tabular}{|c|l|l|l|l|l|}
\hline
$\ell$ & $L$ & $\lambda_1\ell$ & $\lambda_2\ell$ & $\lambda_3\ell$ & $\lambda_4\ell$ \\
\hline
13 &16& 1.878(11) & 2.222(9) & 2.504(7) & 2.737(6)\\
 &20& 1.885(7) & 2.239(6) & 2.520(6) & 2.754(5)\\
 &24& 1.883(18) & 2.237(17) & 2.516(13) & 2.757(10)\\
 &28& 1.909(16) & 2.250(14) & 2.533(12) & 2.770(9)\\
 &$\infty$& 1.930(31) & 2.287(26) & 2.566(22) & 2.811(17)\\
14 &16& 1.813(8) & 2.161(7) & 2.430(6) & 2.665(5)\\
 &20& 1.819(11) & 2.152(11) & 2.429(8) & 2.666(8)\\
 &24& 1.839(11) & 2.174(11) & 2.452(8) & 2.686(8)\\
 &28& 1.858(15) & 2.182(14) & 2.455(11) & 2.688(10)\\
 &$\infty$& 1.899(24) & 2.196(23) & 2.487(19) & 2.719(16)\\
15 &20& 1.748(11) & 2.082(10) & 2.360(8) & 2.591(8)\\
 &24& 1.758(12) & 2.081(11) & 2.365(9) & 2.602(8)\\
 &28& 1.790(16) & 2.127(14) & 2.393(11) & 2.629(10)\\
 &$\infty$& 1.873(52) & 2.201(46) & 2.459(38) & 2.711(35)\\
16 &20& 1.677(13) & 2.015(10) & 2.290(10) & 2.522(9)\\
 &24& 1.715(11) & 2.044(11) & 2.316(8) & 2.546(8)\\
 &28& 1.729(13) & 2.054(12) & 2.325(13) & 2.557(10)\\
 &$\infty$& 1.868(50) & 2.161(44) & 2.424(43) & 2.651(37)\\
17 &20& 1.621(12) & 1.953(12) & 2.230(9) & 2.461(8)\\
 &24& 1.663(7) & 1.980(6) & 2.249(6) & 2.479(5)\\
 &28& 1.668(14) & 1.992(12) & 2.267(12) & 2.489(10)\\
 &$\infty$& 1.814(49) & 2.098(46) & 2.357(40) & 2.564(33)\\
 & & & & & \\
 & & & & & \\
 & & & & & \\
 & & & & & \\
 & & & & & \\
 & & & & & \\
 & & & & & \\
 & & & & & \\
\hline
\end{tabular}
}
\caption{Measurements for $N=8$.}
\eet{tab8}

\bet[H]
\centering
{\renewcommand{\arraystretch}{0.5}
\begin{tabular}{|c|l|l|l|l|l|}
\hline
$\ell$ & $L$ & $\lambda_1\ell$ & $\lambda_2\ell$ & $\lambda_3\ell$ & $\lambda_4\ell$ \\
\hline
2 &16& 3.739(9) & 3.879(9) & 4.027(9) & 4.133(8)\\
 &20& 3.724(11) & 3.868(11) & 4.016(10) & 4.125(9)\\
 &24& 3.738(23) & 3.877(23) & 4.025(22) & 4.133(21)\\
 &28& 3.728(23) & 3.874(24) & 4.021(21) & 4.127(21)\\
 &$\infty$& 3.705(36) & 3.856(37) & 4.003(34) & 4.114(32)\\
4 &16& 3.302(9) & 3.510(9) & 3.702(8) & 3.852(7)\\
 &20& 3.273(8) & 3.489(8) & 3.681(7) & 3.834(8)\\
 &24& 3.300(13) & 3.514(12) & 3.710(12) & 3.857(11)\\
 &28& 3.270(18) & 3.483(17) & 3.676(15) & 3.828(14)\\
 &$\infty$& 3.243(27) & 3.473(27) & 3.667(24) & 3.820(22)\\
6 &16& 2.940(6) & 3.198(6) & 3.422(6) & 3.599(5)\\
 &20& 2.946(9) & 3.201(8) & 3.423(6) & 3.599(6)\\
 &24& 2.949(11) & 3.209(11) & 3.431(9) & 3.605(9)\\
 &28& 2.947(14) & 3.211(12) & 3.437(11) & 3.616(10)\\
 &$\infty$& 2.963(22) & 3.227(21) & 3.450(18) & 3.626(16)\\
8 &16& 2.693(6) & 2.973(6) & 3.211(6) & 3.403(5)\\
 &20& 2.698(6) & 2.982(6) & 3.219(5) & 3.407(4)\\
 &24& 2.686(10) & 2.975(9) & 3.219(7) & 3.413(7)\\
 &28& 2.668(17) & 2.946(16) & 3.189(12) & 3.387(11)\\
 &$\infty$& 2.673(22) & 2.967(21) & 3.210(18) & 3.407(16)\\
10 &16& 2.497(7) & 2.793(6) & 3.038(6) & 3.238(5)\\
 &20& 2.477(11) & 2.775(10) & 3.021(10) & 3.225(9)\\
 &24& 2.501(12) & 2.805(11) & 3.049(8) & 3.249(7)\\
 &28& 2.500(11) & 2.799(10) & 3.044(8) & 3.245(8)\\
 &$\infty$& 2.498(21) & 2.807(18) & 3.054(15) & 3.257(15)\\
12 &16& 2.367(6) & 2.671(6) & 2.912(5) & 3.111(5)\\
 &20& 2.345(10) & 2.646(9) & 2.893(8) & 3.097(6)\\
 &24& 2.349(15) & 2.651(13) & 2.897(10) & 3.098(9)\\
 &28& 2.354(12) & 2.653(12) & 2.903(9) & 3.103(8)\\
 &$\infty$& 2.321(22) & 2.613(21) & 2.877(17) & 3.082(15)\\
\hline
\end{tabular}
}
{\renewcommand{\arraystretch}{0.5}
\begin{tabular}{|c|l|l|l|l|l|}
\hline
$\ell$ & $L$ & $\lambda_1\ell$ & $\lambda_2\ell$ & $\lambda_3\ell$ & $\lambda_4\ell$ \\
\hline
13 &16& 2.291(8) & 2.588(7) & 2.829(6) & 3.029(6)\\
 &20& 2.282(6) & 2.590(5) & 2.833(4) & 3.034(3)\\
 &24& 2.265(10) & 2.577(9) & 2.820(7) & 3.028(7)\\
 &28& 2.284(8) & 2.581(8) & 2.825(8) & 3.030(7)\\
 &$\infty$& 2.260(19) & 2.568(17) & 2.816(15) & 3.032(14)\\
14 &16& 2.239(6) & 2.532(6) & 2.777(6) & 2.981(4)\\
 &20& 2.214(12) & 2.520(11) & 2.766(9) & 2.973(7)\\
 &24& 2.223(12) & 2.521(11) & 2.770(9) & 2.975(7)\\
 &28& 2.232(11) & 2.532(11) & 2.775(8) & 2.980(7)\\
 &$\infty$& 2.207(21) & 2.518(20) & 2.765(16) & 2.972(14)\\
15 &20& 2.173(9) & 2.472(9) & 2.722(7) & 2.923(6)\\
 &24& 2.166(10) & 2.469(9) & 2.711(7) & 2.911(5)\\
 &28& 2.167(12) & 2.469(14) & 2.714(11) & 2.917(9)\\
 &$\infty$& 2.150(41) & 2.461(42) & 2.684(34) & 2.886(28)\\
16 &20& 2.126(10) & 2.425(8) & 2.666(8) & 2.868(6)\\
 &24& 2.108(11) & 2.407(9) & 2.654(8) & 2.856(7)\\
 &28& 2.124(9) & 2.422(8) & 2.668(9) & 2.869(7)\\
 &$\infty$& 2.115(37) & 2.407(32) & 2.663(32) & 2.862(27)\\
17 &20& 2.078(10) & 2.375(9) & 2.613(7) & 2.818(6)\\
 &24& 2.082(11) & 2.369(10) & 2.610(7) & 2.814(7)\\
 &28& 2.065(13) & 2.358(13) & 2.600(10) & 2.805(9)\\
 &$\infty$& 2.050(44) & 2.324(41) & 2.576(32) & 2.780(29)\\
 & & & & & \\
 & & & & & \\
 & & & & & \\
 & & & & & \\
 & & & & & \\
 & & & & & \\
 & & & & & \\
 & & & & & \\
\hline
\end{tabular}
}
\caption{Measurements for $N=12$.}
\eet{tab12}

\bibliography{biblio}
\end{document}